%% file: main.tex
\documentclass[10pt,conference]{IEEEtran}
\usepackage{array}
\input{macros}


\begin{document}

%


\title{Interprocedural Semantic Change-Impact Analysis\\using Equivalence Relations
}

\author{
\IEEEauthorblockN{Alex Gyori}
\IEEEauthorblockA{University of Illinois\\
gyori@illinois.edu}
\and
\IEEEauthorblockN{Shuvendu K Lahiri}
\IEEEauthorblockA{Microsoft Research\\
shuvendu@microsoft.com}
\and
\IEEEauthorblockN{Nimrod Partush}
\IEEEauthorblockA{Technion\\
nimi@cs.technion.ac.il}}
%

\maketitle
\begin{abstract}
Change-impact analysis (CIA) is the task of determining the set of
program elements impacted by a program change.  Precise CIA has
great potential to avoid expensive testing and code reviews for 
(parts of) changes that are refactorings (semantics-preserving).  Existing CIA is imprecise
because it is coarse-grained, deals with only few refactoring
patterns, or is unaware of the change semantics.

We formalize the notion of change impact in terms of the trace
semantics of two program versions.  We show how to leverage
equivalence relations to make dataflow-based CIA aware of the
change semantics, thereby improving precision in the presence of
semantics-preserving changes.  We propose an {\it anytime} algorithm
that allows applying costly equivalence relation inference
incrementally to refine the set of impacted statements.  We have
implemented a prototype in \SymDiff, and evaluated it on \numChanges{}
real-world changes from open-source projects and benchmark programs
used by prior research.  The evaluation results show an average
\avgRed{} improvement in the size of the set of impacted statements
compared to standard dataflow-based techniques.
\end{abstract}




\input{intro}
\input{background}

\input{change}
\input{technique}

\input{evaluation}
\input{related}
\input{conclusions}

\bibliographystyle{abbrv}
\bibliography{biblio}


\end{document}

%% file: macros.tex
\usepackage{booktabs}
\usepackage{cite}
\usepackage{color}
\usepackage{xcolor}
\usepackage{float}
\usepackage{listings}
\usepackage{hyperref}
\usepackage{amssymb}
\usepackage{amsmath}
\usepackage{xspace}
\usepackage{graphicx}
\usepackage{multirow}
\usepackage{url}
\usepackage[ruled,vlined,linesnumbered]{algorithm2e}
\usepackage{mathpartir}
\usepackage[T1]{fontenc}
\usepackage{beramono}
\usepackage{tikz}

\newtheorem{theorem}{Theorem}

\newtheorem{definition}{Definition}

\newcommand{\SymDiff}{\textsc{SymDiff}}

\newcommand{\fig}{Figure}
\newcommand{\tab}{Table}

\newcommand{\CodeIn}[1]{{\small \texttt{#1}}}
\newcommand{\Comment}[1]{}
\newcommand{\Space}[1]{}
\newcommand{\Fix}[1]{\textcolor{red}{[#1]}}

\newcommand{\codein}[1]{\CodeIn{#1}}

\newcommand{\na}{\textit{n.a.}}
\newcommand{\timeout}{\textit{timeout}}

\newcommand{\Def}[2]{\expandafter\newcommand\csname rmk-#1\endcsname{#2}}
\newcommand{\Use}[1]{\csname rmk-#1\endcsname}

\newcommand{\cia}{CIA}
\newcommand{\numChanges}{322}
\newcommand{\numGitChanges}{164}
\newcommand{\numSiemChanges}{158}
\newcommand{\numGithub}{5}
\newcommand{\numBench}{6}

\newcommand{\semimp}{\semDataflowcia}

\definecolor{light}{gray}{0.80}
\newcommand{\highlight}[1]{
  \addtolength{\fboxsep}{-1mm}
  \colorbox{light}{$#1$}
  \addtolength{\fboxsep}{1mm}}

\newcommand{\Programs}{\textit{Programs}}
\newcommand{\programn}{\textit{Prog}}
\newcommand{\program}[1]{\programn^{#1}}
\newcommand{\pprogram}{\program{}} 
\newcommand{\Procs}{\textit{Procs}}
\newcommand{\proc}[1]{{\textit{f}_{#1}}}
\newcommand{\altproc}[1]{\textit{g}_{#1}} 
\newcommand{\pproc}{\proc{}} 
\newcommand{\aaltproc}{\altproc{}} 
\newcommand{\hproc}{\textit{h}}
\newcommand{\mainProc}{\textit{main}} 
\newcommand{\Stmts}{\textit{Stmts}}
\newcommand{\stmtMap}{\textit{StmtAt}}
\newcommand{\stmt}[1]{\textit{st}_{#1}}
\newcommand{\sstmt}{\stmt{}} 

\newcommand{\ProcNodes}[1]{\textit{N}_{#1}}

\newcommand{\procNode}[1]{\textit{n}_{#1}}

\newcommand{\pprocNode}{\procNode{}} 
\newcommand{\procEntryNode}[1]{\procNode{#1}^e}
\newcommand{\procExitNode}[1]{\procNode{#1}^x}
\newcommand{\ProcEdges}[1]{\textit{E}_{#1}}

\newcommand{\ProcIns}[1]{\textit{In}_{#1}}

\newcommand{\ProcOuts}[1]{\textit{Out}_{#1}}

\newcommand{\ProcVars}[1]{\textit{Vars}_{#1}}

\newcommand{\Exprs}{\textit{Exprs}}

\newcommand{\Vars}{\textit{Vars}}

\newcommand{\bfont}[1]{\mathbf{#1}}

\newcommand{\bvar}[1]{\textit{#1}}
\newcommand{\varn}[1]{\bvar{#1}}

\newcommand{\btrue}{\bfont{true}}
\newcommand{\bfalse}{\bfont{false}}

\newcommand{\bop}[1]{\bfont{op}(#1)}

\newcommand{\bskip}{\bfont{skip}}
\newcommand{\bcalln}{\bfont{call}}
\newcommand{\bcall}[3]{{\bcalln} \ {#3} := #1(#2)}

\newcommand{\bifelse}[3]{\bfont{if} \ (#1) \  \ #2 \  \ \bfont{else} \  \ #3 \ }

\newcommand{\bassertn}{\bfont{assert}}
\newcommand{\bassumen}{\bfont{assume}}

\newcommand{\bassign}[2]{{#1} := {#2}}

\newcommand{\bassert}[1]{\bassertn \ #1}
\newcommand{\bassume}[1]{\bassumen \ #1}

\newcommand{\Goto}[1]{\bfont{goto} \ #1}

\newcommand{\seln}{\textit{sel}}
\newcommand{\updaten}{\textit{update}}
\newcommand{\selFunc}[2]{\seln(#1, #2)}
\newcommand{\updateFunc}[3]{\updaten(#1, #2, #3)}

\newcommand{\avgRed}[0]{35\%}

\lstdefinelanguage{boogie}{
  keywords={%
    proc,free, function,returns,var,int,bool,call,return,assert,assume,goto,havoc,modifies,requires,ensures,while,if,const,axiom,foreach
  },
  morecomment=[l]{//},
  basicstyle=\ttfamily\footnotesize,
  escapechar=\%
}

\definecolor{impacted}{rgb}{0.7,0.7,0.7}
\definecolor{syntchanged}{rgb}{1.,0.75,0.75}
\definecolor{linenumbergray}{rgb}{0.5,0.5,0.5}
\definecolor{gitdel}{RGB}{255,236,236}
\definecolor{gitdelfocused}{RGB}{248,203,203}
\definecolor{gitadd}{RGB}{234,255,234}
\definecolor{gitaddfocused}{RGB}{166,235,166}

\makeatletter
\newenvironment{btHighlight}[1][]
{\begingroup\tikzset{bt@Highlight@par/.style={#1}}\begin{lrbox}{\@tempboxa}}
{\end{lrbox}\bt@HL@box[bt@Highlight@par]{\@tempboxa}{impacted}\endgroup}

\newcommand\btHL[1][]{%
  \begin{btHighlight}[#1]\bgroup\aftergroup\bt@HL@endenv%
}
\def\bt@HL@endenv{%
  \end{btHighlight}%
  \egroup
}
\newcommand{\bt@HL@box}[3][]{%
  \tikz[#1]{%
    \pgfpathrectangle{\pgfpoint{1pt}{0pt}}{\pgfpoint{\wd #2}{\ht #2}}%
    \pgfusepath{use as bounding box}%
    \node[anchor=base west, fill=#3, outer sep=0pt,inner xsep=0pt, inner ysep=-0.5pt, #1]{\raisebox{1pt}{\strut}\strut\usebox{#2}};
  }%
}
\makeatother

\lstdefinestyle{C-github}{basicstyle=\ttfamily\scriptsize,
        language=C,
        numbers=left,
        basicstyle=\ttfamily\scriptsize,
        numberstyle=\tiny\color{linenumbergray},
        moredelim=**[is][\btHL]{@i@}{@i@},
        moredelim=**[is][{\btHL[fill=gitdel]}]{@d@}{@d@},
        moredelim=**[is][{\btHL[fill=gitdelfocused, rounded corners=2pt]}]{@df@}{@df@},
        moredelim=**[is][{\btHL[fill=gitadd]}]{@a@}{@a@},
        moredelim=**[is][{\btHL[fill=gitaddfocused, rounded corners=2pt]}]{@af@}{@af@},
        escapeinside={(*@}{@*)}}

\input{tables/nums_results_dac_linear}

\input{tables/nums_projects_summary}

\newcommand{\equalOn}[3]{#1 =_{#3} #2}
\newcommand{\dependencyOf}[1]{\textit{Deps}(#1)}

\newcommand{\ivars}{\textit{impVars}}
\newcommand{\iNodes}{\textit{impNds}}
\newcommand{\isumms}{\textit{impSumms}}
\newcommand{\eqs}{\textit{EQ}}
\newcommand{\preqs}{\textit{prEQ}}

\newcommand{\summeqs}{\textit{smEQ}}

\SetKwFunction{ProcsWithin}{ProcsWithin}
\SetKwFunction{DropProcs}{AbstractProcs}
\newcommand{\kChangedProcs}{\Procs'}

\newcommand{\map}{\pi}
\newcommand{\vmap}{\varpi}
\newcommand{\invMap}{\map^{-1}}
\newcommand{\impacted}[4]{\textit{Impacted}(#1, #2, #3, #4)}
\newcommand{\mappedNodes}{\textsc{mapped}}
\newcommand{\notmapped}[1]{#1 \not\in \mappedNodes}
\newcommand{\taintedNoden}{\textsc{impactedNode}}
\newcommand{\taintedVarn}{\textsc{impactedVar}}
\newcommand{\taintedNode}[1]{\taintedNoden(#1)}
\newcommand{\taintedVar}[1]{\taintedVarn(#1)}
\newcommand{\conditionalNode}[1]{\textsc{branchingNode}(#1)}
\newcommand{\controlDep}[2]{\textsc{ControlDependent}(#2, #1)}
\newcommand{\taintedExpr}[1]{\textsc{ImpactedExpr}(#1)}
\newcommand{\inParam}[3]{\textsc{InFormal}(#1, #2, #3)}
\newcommand{\outParam}[3]{\textsc{OutFormal}(#1, #2, #3)}
\newcommand{\inActual}[4]{\textsc{InActual}(#1, #2, #3, #4)}
\newcommand{\outActual}[4]{\textsc{OutActual}(#1, #2, #3, #4)}
\newcommand{\dependsOnVarn}{\textsc{DependsOnVar}}
\newcommand{\dependsOnNoden}{\textsc{DependsOnNode}}
\newcommand{\dependsOnVar}[3]{\dependsOnVarn(#1, #2, #3)}
\newcommand{\dependsOnNode}[3]{\dependsOnNoden(#1, #2, #3)}
\newcommand{\callsite}[3]{\textsc{Callsite}(#1, #2, #3)}
\newcommand{\taintedSummn}{\textsc{ImpactedSumm}}
\newcommand{\taintedSumm}[2]{\taintedSummn(#1, #2)}
\newcommand{\preEqn}{\textsc{PreEquiv}}

\newcommand{\summEqn}{\textsc{SummaryEquiv}}
\newcommand{\preEq}[2]{\preEqn(#1,#2)}

\newcommand{\summEq}[2]{\summEqn(#1,#2)}

\newcommand{\readSet}[1]{\textit{RVars}({#1})}
\newcommand{\writeSet}[1]{\textit{WVars}({#1})}
\newcommand{\vect}[1]{\overline{#1}}

\newcommand{\val}[1]{\nu_{#1}}
\newcommand{\Values}{{\mathcal V}}
\newcommand{\botValue}{{\bot}}
\newcommand{\store}[1]{{\theta}_{#1}}
\newcommand{\sstore}{{\store{}}}
\newcommand{\Stores}{{\Theta}}
\newcommand{\storeAt}[2]{\store{#1}({#2})}
\newcommand{\sstoreAt}[1]{\storeAt{}{#1}}
\newcommand{\storeUnion}[2]{{#1} \oplus {#2}}
\newcommand{\mkStore}[1]{[{#1}]}
\newcommand{\projStore}[2]{{#1}\vert_{#2}}

\newcommand{\States}{\Sigma}
\newcommand{\state}[1]{\sigma_{#1}}
\newcommand{\sstate}{\state{}} 

\newcommand {\Cstcks}[1]{\textit{CS}_{#1}}
\newcommand {\cstck}[1]{\textit{cs}_{#1}} 
\newcommand {\ccstck}{{\cstck{}}}
\newcommand {\emptyStck}{{\epsilon}}
\newcommand {\concatStck}[2]{#1 :: #2}

\newcommand{\step}[2]{{#1} \leadsto {#2}}
\newcommand{\stepTransitive}[2]{{#1} \leadsto^* {#2}}
\newcommand{\transRel}[1]{\Omega_{#1}}

\newcommand{\trace}[1]{\tau_{#1}}
\newcommand{\traceAt}[2]{\trace{#1}[#2]}
\newcommand{\ProcTraces}[1]{\Gamma_{#1}}
\newcommand{\ProcTracesStarting}[3]{\trace{#1}^{#3}(#2)}
\newcommand{\projectTraceOnNode}[2]{{#1}\vert_{#2}}
\newcommand{\varSeqInStateSeq}[2]{{#1} \rfloor_{#2}}

\newcommand\pfun{\mathrel{\ooalign{\hfil$\mapstochar\mkern5mu$\hfil\cr$\to$\cr}}}

\newcommand{\prodProg}[2]{\program{{#1} \times {#2}}}

\newcommand{\prodProc}[3]{{#1}^{{#2}\times{#3}}}
\newcommand{\takenProc}[1]{\textit{callHist}^{#1}}

\newcommand{\myparabf}[1]{\noindent\textbf{#1:}}
\newcommand{\spcing}{-0.1em}
\newcommand{\vsp}{-1em}

%% file: tables/nums_results_dac_linear.tex
\Def{0.dcf.min.flingfd}{64}
\Def{0.dcf.max.flingfd}{84}
\Def{0.sem.min.flingfd}{39}
\Def{0.sem.max.flingfd}{83}
\Def{0.dcf.time.flingfd}{0.94}
\Def{0.sem.time.flingfd}{8.92}

\Def{0.dcf.min.histo}{0}
\Def{0.dcf.max.histo}{86}
\Def{0.sem.min.histo}{0}
\Def{0.sem.max.histo}{75}
\Def{0.dcf.time.histo}{2.14}
\Def{0.sem.time.histo}{19.43}

\Def{0.dcf.min.mdp}{0}
\Def{0.dcf.max.mdp}{465}
\Def{0.sem.min.mdp}{0}
\Def{0.sem.max.mdp}{330}
\Def{0.dcf.time.mdp}{28.16}
\Def{0.sem.time.mdp}{77.71}

\Def{0.dcf.min.tinyvm}{0}
\Def{0.dcf.max.tinyvm}{344}
\Def{0.sem.min.tinyvm}{0}
\Def{0.sem.max.tinyvm}{308}
\Def{0.dcf.time.tinyvm}{68.96}
\Def{0.sem.time.tinyvm}{158.33}

\Def{1.dcf.min.flingfd}{64}
\Def{1.dcf.max.flingfd}{84}
\Def{1.sem.min.flingfd}{14}
\Def{1.sem.max.flingfd}{70}
\Def{1.dcf.time.flingfd}{0.90}
\Def{1.sem.time.flingfd}{9.85}

\Def{1.dcf.min.histo}{0}
\Def{1.dcf.max.histo}{86}
\Def{1.sem.min.histo}{0}
\Def{1.sem.max.histo}{65}
\Def{1.dcf.time.histo}{2.14}
\Def{1.sem.time.histo}{20.59}

\Def{1.dcf.min.mdp}{0}
\Def{1.dcf.max.mdp}{465}
\Def{1.sem.min.mdp}{0}
\Def{1.sem.max.mdp}{324}
\Def{1.dcf.time.mdp}{28.20}
\Def{1.sem.time.mdp}{100.68}

\Def{1.dcf.min.tinyvm}{0}
\Def{1.dcf.max.tinyvm}{344}
\Def{1.sem.min.tinyvm}{0}
\Def{1.sem.max.tinyvm}{298}
\Def{1.dcf.time.tinyvm}{69.13}
\Def{1.sem.time.tinyvm}{160.35}

\Def{inf.dcf.min.flingfd}{64}
\Def{inf.dcf.max.flingfd}{84}
\Def{inf.sem.min.flingfd}{14}
\Def{inf.sem.max.flingfd}{70}
\Def{inf.dcf.time.flingfd}{0.89}
\Def{inf.sem.time.flingfd}{10.44}

\Def{inf.dcf.min.histo}{0}
\Def{inf.dcf.max.histo}{86}
\Def{inf.sem.min.histo}{0}
\Def{inf.sem.max.histo}{65}
\Def{inf.dcf.time.histo}{2.14}
\Def{inf.sem.time.histo}{24.92}

\Def{inf.dcf.min.mdp}{0}
\Def{inf.dcf.max.mdp}{465}
\Def{inf.sem.min.mdp}{0}
\Def{inf.sem.max.mdp}{283}
\Def{inf.dcf.time.mdp}{29.22}
\Def{inf.sem.time.mdp}{173.08}

\Def{inf.dcf.min.tinyvm}{0}
\Def{inf.dcf.max.tinyvm}{344}
\Def{inf.sem.min.tinyvm}{0}
\Def{inf.sem.max.tinyvm}{283}
\Def{inf.dcf.time.tinyvm}{69.43}
\Def{inf.sem.time.tinyvm}{169.05}

\Def{0.avg.red.flingfd}{20.1\%}
\Def{0.avg.red.histo}{11.5\%}
\Def{0.avg.red.mdp}{1.5\%}
\Def{0.avg.red.tinyvm}{18.5\%}

\Def{1.avg.red.flingfd}{47.3\%}
\Def{1.avg.red.histo}{28.6\%}
\Def{1.avg.red.mdp}{3.4\%}
\Def{1.avg.red.tinyvm}{23.2\%}

\Def{inf.avg.red.flingfd}{47.3\%}
\Def{inf.avg.red.histo}{28.6\%}
\Def{inf.avg.red.mdp}{6.5\%}
\Def{inf.avg.red.tinyvm}{43.6\%}

\Def{1.sem.min.theft}{11}
\Def{1.sem.max.theft}{185}
\Def{1.sem.time.theft}{57.48}
\Def{1.dcf.min.theft}{184}
\Def{1.dcf.max.theft}{261}
\Def{1.dcf.time.theft}{4.52}
\Def{1.avg.red.theft}{62\%}
\Def{inf.sem.min.theft}{11}
\Def{inf.sem.max.theft}{107}
\Def{inf.sem.time.theft}{289.75}
\Def{inf.dcf.min.theft}{184}
\Def{inf.dcf.max.theft}{261}
\Def{inf.dcf.time.theft}{4.43}
\Def{inf.avg.red.theft}{77\%}
\Def{0.sem.min.theft}{11}
\Def{0.sem.max.theft}{186}
\Def{0.sem.time.theft}{38.45}
\Def{0.dcf.min.theft}{184}
\Def{0.dcf.max.theft}{261}
\Def{0.dcf.time.theft}{4.48}
\Def{0.avg.red.theft}{61\%}

\Def{inf.sem.min.space}{\na}
\Def{inf.sem.max.space}{\na}
\Def{inf.avg.red.space}{\na}
\Def{inf.sem.time.space}{\timeout}

\Def{1.sem.min.space}{14}
\Def{1.sem.max.space}{2816}
\Def{1.sem.time.space}{895.14}
\Def{1.dcf.min.space}{20}
\Def{1.dcf.max.space}{2851}
\Def{1.dcf.time.space}{59.30}
\Def{1.avg.red.space}{36.71\%}
\Def{0.sem.min.space}{14}
\Def{0.sem.max.space}{2816}
\Def{0.sem.time.space}{798.96}
\Def{0.dcf.min.space}{20}
\Def{0.dcf.max.space}{2851}
\Def{0.dcf.time.space}{59.45}
\Def{0.avg.red.space}{31.87\%}

\Def{2.sem.min.space}{14}
\Def{2.sem.max.space}{2816}
\Def{2.sem.time.space}{1300.43}
\Def{2.dcf.min.space}{20}
\Def{2.dcf.max.space}{2851}
\Def{2.dcf.time.space}{58.11}
\Def{2.avg.red.space}{40.56\%}
\Def{3.sem.min.space}{14}
\Def{3.sem.max.space}{2808}
\Def{3.sem.time.space}{1900.03}
\Def{3.dcf.min.space}{20}
\Def{3.dcf.max.space}{2851}
\Def{3.dcf.time.space}{58.35}
\Def{3.avg.red.space}{43.96\%}

\Def{1.sem.min.print_tokens}{34}
\Def{1.sem.max.print_tokens}{128}
\Def{1.sem.time.print_tokens}{58.23}
\Def{1.dcf.min.print_tokens}{151}
\Def{1.dcf.max.print_tokens}{153}
\Def{1.dcf.time.print_tokens}{2.22}
\Def{1.avg.red.print_tokens}{28.67\%}
\Def{1.sem.min.replace}{70}
\Def{1.sem.max.replace}{194}
\Def{1.sem.time.replace}{92.72}
\Def{1.dcf.min.replace}{75}
\Def{1.dcf.max.replace}{195}
\Def{1.dcf.time.replace}{4.95}
\Def{1.avg.red.replace}{2.89\%}
\Def{0.sem.min.tcas}{0}
\Def{0.sem.max.tcas}{49}
\Def{0.sem.time.tcas}{7.94}
\Def{0.dcf.min.tcas}{1}
\Def{0.dcf.max.tcas}{49}
\Def{0.dcf.time.tcas}{0.66}
\Def{0.avg.red.tcas}{9.24\%}
\Def{inf.sem.min.print_tokens}{34}
\Def{inf.sem.max.print_tokens}{128}
\Def{inf.sem.time.print_tokens}{102.40}
\Def{inf.dcf.min.print_tokens}{151}
\Def{inf.dcf.max.print_tokens}{153}
\Def{inf.dcf.time.print_tokens}{2.22}
\Def{inf.avg.red.print_tokens}{28.67\%}
\Def{inf.sem.min.replace}{65}
\Def{inf.sem.max.replace}{174}
\Def{inf.sem.time.replace}{236.77}
\Def{inf.dcf.min.replace}{75}
\Def{inf.dcf.max.replace}{195}
\Def{inf.dcf.time.replace}{4.95}
\Def{inf.avg.red.replace}{9.41\%}
\Def{1.sem.min.tot_info}{24}
\Def{1.sem.max.tot_info}{102}
\Def{1.sem.time.tot_info}{74.99}
\Def{1.dcf.min.tot_info}{103}
\Def{1.dcf.max.tot_info}{104}
\Def{1.dcf.time.tot_info}{6.38}
\Def{1.avg.red.tot_info}{46.26\%}
\Def{0.sem.min.print_tokens2}{80}
\Def{0.sem.max.print_tokens2}{129}
\Def{0.sem.time.print_tokens2}{16.08}
\Def{0.dcf.min.print_tokens2}{155}
\Def{0.dcf.max.print_tokens2}{158}
\Def{0.dcf.time.print_tokens2}{1.46}
\Def{0.avg.red.print_tokens2}{30.36\%}
\Def{0.sem.min.tot_info}{31}
\Def{0.sem.max.tot_info}{102}
\Def{0.sem.time.tot_info}{37.01}
\Def{0.dcf.min.tot_info}{103}
\Def{0.dcf.max.tot_info}{104}
\Def{0.dcf.time.tot_info}{6.39}
\Def{0.avg.red.tot_info}{18.65\%}
\Def{1.sem.min.tcas}{0}
\Def{1.sem.max.tcas}{49}
\Def{1.sem.time.tcas}{8.63}
\Def{1.dcf.min.tcas}{1}
\Def{1.dcf.max.tcas}{49}
\Def{1.dcf.time.tcas}{0.66}
\Def{1.avg.red.tcas}{9.24\%}
\Def{0.sem.min.print_tokens}{69}
\Def{0.sem.max.print_tokens}{137}
\Def{0.sem.time.print_tokens}{24.02}
\Def{0.dcf.min.print_tokens}{151}
\Def{0.dcf.max.print_tokens}{153}
\Def{0.dcf.time.print_tokens}{2.22}
\Def{0.avg.red.print_tokens}{19.37\%}
\Def{inf.sem.min.tot_info}{12}
\Def{inf.sem.max.tot_info}{77}
\Def{inf.sem.time.tot_info}{127.61}
\Def{inf.dcf.min.tot_info}{103}
\Def{inf.dcf.max.tot_info}{104}
\Def{inf.dcf.time.tot_info}{6.38}
\Def{inf.avg.red.tot_info}{56.50\%}
\Def{inf.sem.min.schedule}{7}
\Def{inf.sem.max.schedule}{68}
\Def{inf.sem.time.schedule}{30.83}
\Def{inf.dcf.min.schedule}{79}
\Def{inf.dcf.max.schedule}{115}
\Def{inf.dcf.time.schedule}{1.38}
\Def{inf.avg.red.schedule}{70.85\%}
\Def{0.sem.min.replace}{74}
\Def{0.sem.max.replace}{194}
\Def{0.sem.time.replace}{35.72}
\Def{0.dcf.min.replace}{75}
\Def{0.dcf.max.replace}{195}
\Def{0.dcf.time.replace}{4.96}
\Def{0.avg.red.replace}{2.08\%}
\Def{1.sem.min.schedule}{7}
\Def{1.sem.max.schedule}{87}
\Def{1.sem.time.schedule}{24.15}
\Def{1.dcf.min.schedule}{79}
\Def{1.dcf.max.schedule}{115}
\Def{1.dcf.time.schedule}{1.37}
\Def{1.avg.red.schedule}{40.58\%}
\Def{1.sem.min.print_tokens2}{59}
\Def{1.sem.max.print_tokens2}{101}
\Def{1.sem.time.print_tokens2}{24.42}
\Def{1.dcf.min.print_tokens2}{155}
\Def{1.dcf.max.print_tokens2}{158}
\Def{1.dcf.time.print_tokens2}{1.46}
\Def{1.avg.red.print_tokens2}{44.65\%}
\Def{0.sem.min.schedule}{7}
\Def{0.sem.max.schedule}{104}
\Def{0.sem.time.schedule}{13.73}
\Def{0.dcf.min.schedule}{79}
\Def{0.dcf.max.schedule}{115}
\Def{0.dcf.time.schedule}{1.37}
\Def{0.avg.red.schedule}{26.35\%}
\Def{inf.sem.min.print_tokens2}{55}
\Def{inf.sem.max.print_tokens2}{100}
\Def{inf.sem.time.print_tokens2}{97.98}
\Def{inf.dcf.min.print_tokens2}{155}
\Def{inf.dcf.max.print_tokens2}{158}
\Def{inf.dcf.time.print_tokens2}{1.46}
\Def{inf.avg.red.print_tokens2}{45.66\%}
\Def{inf.sem.min.tcas}{0}
\Def{inf.sem.max.tcas}{49}
\Def{inf.sem.time.tcas}{9.71}
\Def{inf.dcf.min.tcas}{1}
\Def{inf.dcf.max.tcas}{49}
\Def{inf.dcf.time.tcas}{0.66}
\Def{inf.avg.red.tcas}{9.24\%}

\Def{0.avg.red}{22\%}
\Def{0.avg.overhead}{9x}
\Def{1.avg.red}{\Fix{?}}
\Def{1.avg.overhead}{\Fix{?}}
\Def{inf.avg.red}{\Fix{?}}
\Def{inf.avg.overhead}{\Fix{?}}

%% file: tables/nums_projects_summary.tex
\Def{pn.tinyvm}{tinyvm}
\Def{vn.tinyvm}{61}
\Def{sloc.min.tinyvm}{425}
\Def{sloc.max.tinyvm}{903}
\Def{cng.min.tinyvm}{1}
\Def{cng.max.tinyvm}{328}
\Def{pn.theft}{theft}
\Def{vn.theft}{2}
\Def{sloc.min.theft}{1672}
\Def{sloc.max.theft}{1838}
\Def{cng.min.theft}{2}
\Def{cng.max.theft}{2}
\Def{pn.histo}{histo}
\Def{vn.histo}{8}
\Def{sloc.min.histo}{617}
\Def{sloc.max.histo}{624}
\Def{cng.min.histo}{1}
\Def{cng.max.histo}{6}
\Def{pn.flingfd}{flingfd}
\Def{vn.flingfd}{2}
\Def{sloc.min.flingfd}{142}
\Def{sloc.max.flingfd}{146}
\Def{cng.min.flingfd}{2}
\Def{cng.max.flingfd}{14}
\Def{pn.mdp}{mdp}
\Def{vn.mdp}{91}
\Def{sloc.min.mdp}{135}
\Def{sloc.max.mdp}{1616}
\Def{cng.min.mdp}{1}
\Def{cng.max.mdp}{402}

\Def{pn.space}{space}
\Def{vn.space}{38}
\Def{sloc.min.space}{6180}
\Def{sloc.max.space}{6205}
\Def{cng.min.space}{1}
\Def{cng.max.space}{42}

\Def{pn.pt}{print\_tokens}
\Def{vn.pt}{5}
\Def{sloc.min.pt}{478}
\Def{sloc.max.pt}{480}
\Def{cng.min.pt}{1}
\Def{cng.max.pt}{8}

\Def{pn.pt2}{print\_tokens2}
\Def{vn.pt2}{10}
\Def{sloc.min.pt2}{397}
\Def{sloc.max.pt2}{402}
\Def{cng.min.pt2}{1}
\Def{cng.max.pt2}{6}

\Def{pn.replace}{replace}
\Def{vn.replace}{32}
\Def{sloc.min.replace}{509}
\Def{sloc.max.replace}{516}
\Def{cng.min.replace}{1}
\Def{cng.max.replace}{15}

\Def{pn.schedule}{schedule}
\Def{vn.schedule}{9}
\Def{sloc.min.schedule}{290}
\Def{sloc.max.schedule}{294}
\Def{cng.min.schedule}{2}
\Def{cng.max.schedule}{4}

\Def{pn.tcas}{tcas}
\Def{vn.tcas}{41}
\Def{sloc.min.tcas}{136}
\Def{sloc.max.tcas}{140}
\Def{cng.min.tcas}{2}
\Def{cng.max.tcas}{16}

\Def{pn.tot_info}{tot\_info}
\Def{vn.tot_info}{23}
\Def{sloc.min.tot_info}{346}
\Def{sloc.max.tot_info}{347}
\Def{cng.min.tot_info}{2}
\Def{cng.max.tot_info}{3}

%% file: intro.tex

\section{Introduction}
\label{sec:introduction}


Software constantly evolves to add and improve features, eliminate bugs, improve software design, etc.
As software evolves faster than ever, it requires rigorous techniques to ensure that changes do not modify existing behavior in unintended ways. 
Several approaches have emerged to address this issue. 
Code review~\cite{mcintosh2014impact}, regression testing~\cite{Rothermel1997SERTS, gligoric2015practical}, test-suite augmentation~\cite{Person2011DISE, Person2008DSE, marinescu2012make}, code contracts~\cite{meyer-contracts, barnett2004spec}, regression verification~\cite{pastore2014verification,godlin-dac09} and verification modulo versions~\cite{vmv:logozzo} are just a few of the approaches to ensure the quality of a change; they all directly benefit from change-impact analysis~(\cia).
 

{\it Change-Impact Analysis} determines the set of program elements that may be impacted by a syntactic change.
Traditional approaches are coarse-grained and operate at the level of types and classes~\cite{apiwattanapong2005icse, apiwattanapong2004differencing}, or files~\cite{gligoric2015practical} to retain soundness. 
Fine-grained techniques that aim to work at the level of statements are typically based on performing dataflow analysis~\cite{ipdfs-cite} on one program to propagate the change along data and control flow edges~\cite{backes2013regression,lehnert2011review, haipeng2015survey}.
Such techniques fail to take the {\it semantics of the change} into account; therefore, they cannot distinguish between changes that a user expects to have only local impact on existing code (e.g., a code refactoring) from ones that have substantial impact on existing code (e.g., changing the functionality or fixing a bug).
The ability to distinguish changes whose impact is local (limited to the changed procedure or a few callers or callees within one or two levels) can help with code review and regression-testing efforts. 
Changes with substantial impact can be prioritized for more rigorous code reviews and may require more testing.

In this paper, we aim to improve the precision of CIA by leveraging \emph{equivalence relations} between variables of two programs across a change. 
At a high level, these equivalences help prune the flow of a change along the data or control flow edges of the changed program. 
To integrate such equivalences, we first formalize the notion of change impact precisely in terms of the trace semantics of two programs.
Next, we show how to make CIA change-semantics aware by incorporating various equivalence relations into an interprocedural dataflow analysis. 
Since computing equivalence relations is expensive, we propose an {\it anytime algorithm}~\cite{anytime-wiki,zilberstein1995approximate} to incrementally compute equivalence relations.

\input{overview2.tex}

\subsection{Contributions}
In this work, we make the following contributions:
\begin{enumerate}
\setlength\itemsep{\spcing}
\item We precisely formalize the set of statements {\it impacted} by a change, in terms of the trace semantics of two programs (\S~\ref{sec:change}).
\item We make a dataflow-based CIA change-semantics aware by incorporating various equivalence relations (\S~\ref{sec:equiv}).
\item We describe an anytime algorithm that allows incrementally computing more equivalences to refine the analysis at the expense of time~(\S~\ref{sec:anytime}).
\item  We have implemented a prototype using \SymDiff~\cite{lahiri-cav12,lahiri-fse13}, and evaluated our technique on \numChanges{} real-world changes collected from GitHub open-source projects and several standard benchmark programs used in prior research~\cite{hutchins1994experiments}.
\end{enumerate}

\myparabf{Verifiability}
The prototype implementation of our tool is open source and available in the SymDiff repository \url{https://symdiff.codeplex.com/}. 
The usage options of the tool are described at \url{https://symdiff.codeplex.com/wikipage?title=Change-Impact}. 
Details about our experimental subjects are available at \url{http://mir.cs.illinois.edu/gyori/projects/cia.html}

%% file: overview2.tex

\subsection{Overview}
\label{sec:overview}

\begin{figure}[h!]
\begin{lstlisting}[style=C-github]
@a@+ static unsigned char line_delim = '\n';@a@ (*@\label{line:change10}@*)
int print_product_info(int name, int version) {
  int locale, printed = 0;
@d@- print_header(@df@'\n'@df@);@d@ (*@\label{line:change11}@*)
@a@+ print_header(@af@line_delim@af@);@a@ // spurious arg impact (*@\label{line:change12}@*)
  locale = locale_ok(); // spurious impact (*@\label{line:localeok}@*)
  if (name) {
    printed = print_name(locale);  (*@\label{line:printname}@*)
  }
  if (version && printed) {
    printed = print_major_version(locale));
@d@-   printed = print_minor_version(locale,@df@'\n'@df@);@d@(*@\label{line:change22}@*)
@a@+   line_delim = '\0';@a@ (*@\label{line:change20}@*)
@a@+   printed = print_minor_version(locale,@af@line_delim@af@));@a@ (*@\label{line:change21}@*)
  }
  return printed;
}
void print_header(char delim) {
  printf("%s%c",HEADER,delim);
}
int locale_ok() {
@d@- return setlocale (LC_ALL,"") @df@? 1 : 0@df@;@d@ (*@\label{line:change30}@*)
@a@+ return @af@!!@af@setlocale (LC_ALL,"");@a@ (*@\label{line:change31}@*)
}
int print_name(int locale) {
  if (locale) {
    printf("%s",locale_format("Coreutils"));
    return 1;
  }
  return 0;
}
int print_major_version(int locale) {
  if (locale) {
    printf("%s",locale_format("8"));
    return 1;
  }
  return 0;
}
int print_minor_version(int locale, char delim) {
  if (locale) {
@i@    printf("%s%c",locale_format(".12"),delim);@i@ (*@\label{line:impacted}@*)
    return 1;
  }
  return 0;
}
\end{lstlisting}
\vspace{\vsp}
\caption{Example. The lines with $-$ and $+$ represent deleted and added lines, respectively.}
\label{fig:running-example}
\vspace{\vsp}
\vspace{\vsp}
\end{figure}
 
Figure~\ref{fig:running-example} shows a running example written in \textsc{C}\Comment{ with several procedures}.
The example is inspired by real commits to \texttt{Coreutils}, in files \texttt{paste.c}~\cite{coreutils-paste-commit} and \texttt{sort.c}~\cite{coreutils-sort-commit}. 
The program has three changes. Two are semantics-preserving changes: 
(i) extracting the literal \texttt{'$\backslash$n'} into the variable \texttt{line\_delim} in the procedure \texttt{print\_product\_info} (lines \ref{line:change10}, \ref{line:change11}, \ref{line:change12}) and 
(ii) replacing the conditional operator with a double negation in \texttt{locale\_ok} (lines \ref{line:change30}, \ref{line:change31})\footnote{Negation in C coerces the values to 0 or 1.}.
The third change sets the \texttt{line\_delim} variable to \texttt{'$\backslash$0'} in the procedure \texttt{print\_product\_info} (lines \ref{line:change22}, \ref{line:change20}, \ref{line:change21}), which impacts statements in \texttt{print\_minor\_version}.
Assume for this example that all executions start from \texttt{print\_product\_info}.
We claim that the only (syntactically unchanged)  line that is {\it impacted by the changes} is the highlighted line~\ref{line:impacted} due to the semantic change at the callsite; 
a statement is impacted, intuitively, if the \emph{sequence of values} it reads can differ when executing the two versions of the program in the same environment.
For brevity, the example does not contain the definitions of the \texttt{setlocale} and \texttt{locale\_format} procedures as well as the \texttt{LC\_ALL} and \texttt{HEADER} constants as they are not impacted nor relevant. 
We will now analyze the change through the lens of a standard dataflow analysis~\cite{ipdfs-cite} and traditional equivalence checking~\cite{godlin-dac09,lahiri-cav12} and then sketch our technique.

\myparabf{Dataflow}
A dataflow analysis technique starts at the sources of change and propagates them through data and control edges (typically in the changed program). 
Dataflow techniques are not aware of the change semantics, and thus cannot exploit semantics-preserving changes. 
Initially, the call to \texttt{print\_header} on line~\ref{line:change12} has a change to its argument that marks all the statements in the procedure as impacted because they all depend on the changed argument. 
Next, the call to \texttt{locale\_ok} on line~\ref{line:localeok}  impacts the \texttt{locale} variable because of the change to the body of \texttt{locale\_ok} and the data dependency of the return value on the change.
This in turn will mark the input of \texttt{print\_name} as impacted at line~\ref{line:printname}, which in turn flows to its output because the return value is control dependent on the input variable marked as impacted (a context-insensitive analysis will impact the return at {\it all} call sites to \texttt{print\_name}).
This impact through the return value will propagate to the call to \texttt{print\_major\_version} and \texttt{print\_minor\_version} because of the control dependency on \texttt{printed} and will impact all the statements in these procedures as well as all the returns at both call sites. 
Finally, the call to \texttt{print\_minor\_version} will impact all of the callee statements. 
A context-sensitive analysis does not help either because the body of \texttt{locale\_ok} changes, which implies that the return value may change across the two versions. 
This is {\it imprecise} since the analysis is unable to determine that the statements in \texttt{print\_name} and \texttt{print\_major\_version} are not impacted.

\myparabf{Equivalence} A traditional interprocedural equivalence checking~\cite{godlin-dac09,lahiri-cav12} (that checks if two procedures have identical input-output behavior) will find that \texttt{locale\_ok}, \texttt{print\_name}, \texttt{print\_header}, \texttt{print\_major\_version}, and \texttt{print\_minor\_version} have identical summaries. This is {\it unsound} for the question of impact analysis, as the statement of \texttt{print\_minor\_version} is impacted due to the change of print delimiter. This illustrates the difference between CIA and (traditional) equivalence checking: two procedures can be equivalent, but still impacted, because  they may get called under different contexts and exhibit different behaviors.

\myparabf{Our approach} In this work, we present a {\it change-semantics aware} CIA that works as follows:
it infers equivalence relations over variables and determines that the arguments at all call sites to \texttt{print\_name} and \texttt{print\_major\_version} are equal across both versions and stops propagating impacts through their arguments.
Further, \texttt{locale\_ok} has an equivalent summary in the two versions (by using equivalence checking)---this ensures that two call sites with equal arguments return equal results.
From these two facts, the technique infers (by simple dataflow) that arguments to \texttt{print\_name} and \texttt{print\_major\_version} are not impacted and therefore the statements in \texttt{print\_name} and \texttt{print\_major\_version} are not impacted.
Thus, our approach precisely identifies the only unchanged impacted line as line~\ref{line:impacted}.

%% file: background.tex
\section{Background}

\input{language}

%% file: language.tex

For the rest of the paper, we will formalize the problem and our technique over a simple language.
We can compile most features of general purpose imperative programming languages to our simple language; we discuss this in \S~\ref{sec:expres}.

\subsection{A Simple Language}
\label{sec:language}

%
%

A program consists of procedures represented as control-flow graphs, statements, and expressions.

\myparabf{Expressions} $e \in \Exprs$ in the language are built up from constants, variables and operator applications:
\[
\begin{array}{lll}
e \in \Exprs & :: & c \ | \  \varn{x} \ | \  \varn{y} \ | \ \ldots \ | \  \bop{e_1,\ldots,e_k} 
\end{array}
\]
Here $c$ represents {\it constant} values of different types such as $\{\btrue, \bfalse\}$ for Booleans, $\{\ldots, -1, 0, 1, \ldots\}$ for integers, and 
$\varn{x}$ denotes variables in scope.
An operator $\bfont{op}$ is a function or predicate symbol that can be uninterpreted or interpreted by some theories (e.g., $\{+, -, *, \leq, \geq, \ldots\}$ by the theory of  arithmetic).
We represent a vector of variables and expressions using $\vect{\varn{x}}$ and $\vect{e}$, respectively.

\myparabf{Statements} $\sstmt \in \Stmts$ are comprised of {\it assign}, {\it assume}, {\it skip} and procedure {\it call} statements.
\[
\begin{array}{lll}
\sstmt \in \Stmts &::& \bassign{\varn{x}}{e} \ | \ \bassume{e} \ | \ \bskip \ | \  \\
                  &  & \bcall{\pproc}{e_1, e_2, \ldots, e_m}{\varn{x}_1, \varn{x}_2, \ldots, \varn{x}_k}
\end{array}
\]
The argument to $\bassumen$ is a Boolean-valued expression, and a $\bskip$ is a no-op.
A call statement can have multiple return values and they are assigned to variables $\varn{x}_i$ at the call site. 

\myparabf{Procedures} A procedure $\pproc{} \in \Procs$ is represented as a control-flow graph consisting of $(\ProcNodes{\pproc}, \ProcEdges{\pproc}, \ProcIns{\pproc}, \ProcOuts{\pproc}, \ProcVars{\pproc}, \procEntryNode{\pproc}, \procExitNode{\pproc})$, where:
\begin{itemize}	
\setlength\itemsep{\spcing}
\item $\ProcNodes{\pproc}$ is a set of control-flow locations in $\pproc$,
\item $\ProcEdges{\pproc} \subseteq \ProcNodes{\pproc} \times \ProcNodes{\pproc}$ is a set of edges over $\ProcNodes{\pproc}$ denoting control-flow,
\item $\ProcIns{\pproc}$ (respectively, $\ProcOuts{\pproc}$) is the vector of {\it input} (respectively, {\it output}) formals of $\pproc$. The output formals model return values and out parameters. 
\item $\ProcVars{\pproc}$ is the set of variables in the scope of $\pproc$ and includes $\ProcIns{\pproc}$, $\ProcOuts{\pproc}$, and local variables of $\pproc$,
\item $\procEntryNode{\pproc} \in \ProcNodes{\pproc}$ (respectively, $\procExitNode{\pproc} \in \ProcNodes{\pproc}$) is the unique {\it entry} (respectively, {\it exit}) node of $\pproc$.
\end{itemize}

Let $\ProcNodes{} = \bigcup_{\pproc \in \Procs} \ProcNodes{\pproc}$ and $\ProcVars{} = \bigcup_{\pproc \in \Procs} \ProcVars{\pproc}$.
Nodes and variables in a procedure $\pproc$ are often denoted by $\procNode{\pproc}$ and $\varn{x}_{\pproc}$ respectively.
For any node $\procNode{\pproc} \in \ProcNodes{\pproc}$, we define the {\it readset} $\readSet{\procNode{\pproc}}$ and {\it writeset} $\writeSet{\procNode{\pproc}}$ as the set of variables that are read and written to respectively in the statement at $\procNode{\pproc}$.

A {\it program} $\pprogram \in \Programs$ is a tuple $(\Procs, \mainProc, \stmtMap)$
where (i) $\Procs$ is  a set of procedures in the program, 
(ii) $\mainProc \in \Procs$ is  the entry procedure from which the program execution starts, and 
(iii) $\stmtMap : \ProcNodes{} \rightarrow \Stmts$ maps a node $\pprocNode \in \ProcNodes{}$ in a procedure $\pproc$ to a {\it statement}. 
For any $\pproc$, we assume that $\stmtMap(\procExitNode{\pproc}) = \bskip$. 

\subsection{Expressiveness}
\label{sec:expres}
We can compile most constructs in general purpose imperative programming languages to our simple language.
This follows the same principle as translators from languages such as C and Java to the Boogie language~\cite{boogie,smack,havoc,flanagan-pldi02}.

\myparabf{Control flow}
Loops can be automatically transformed into tail-recursive procedures~\cite{lahiri-fse13,lahiri-cav12,godlin-dac09}.
We use  $\procNode{1}: \sstmt; \Goto{\procNode{2}, \procNode{3}};$
to express that $\stmtMap(\procNode{1}) = \sstmt$ and $\{(\procNode{1},\procNode{2}) (\procNode{1},\procNode{3})\} \subseteq \ProcEdges{\pproc}$.
A conditional statement $\bifelse{e}{\stmt{1}}{\stmt{2}}$ is modeled as: 
\[
\begin{array}{c}
\procNode{1}: \bassign{\varn{x}}{e};  \Goto{\procNode{2}, \procNode{3}}; \\
\procNode{2}:  \bassume{\varn{x}}; \stmt{1}; \Goto{\procNode{4}}; 
\procNode{3}:  \bassume{\neg \varn{x}}; \stmt{2}; \Goto{\procNode{4}};
\end{array}
\]
where a fresh Boolean variable $\varn{x}$ captures the value of the condition $e$\footnote{The introduction of $\varn{x}$ simplifies determining if control flow is impacted by only inspecting the conditional node}. 
We assume that each node $\pprocNode \in \ProcNodes{\pproc}$ has at most two successor nodes in $\ProcEdges{\pproc}$, where nodes with two successors correspond to conditional statements. 
The only use of an $\bassumen$ statement is to model a conditional statement. 
We refer to $\procNode{1}$ as a {\it branching} node with two successors in $\ProcEdges{}$ with complementary expressions in $\bassumen$ statements.

\myparabf{Globals and heap}~Richer data types such as arrays and maps can be modeled, e.g.,  array read $\varn{x}[e]$ is modeled using $\selFunc{\varn{x}}{e}$ and a write $\varn{x}[e_1] := e_2$ is modeled using $\varn{x} := \updateFunc{\varn{x}}{e_1}{e_2}$~\cite{smt-lib}.
Arrays are in turn used to model the heap in imperative programs and are standard in most software verification tools~\cite{flanagan-pldi02,havoc,smack}.
Additional internal non-determinism (e.g. read from file, network) is lifted as reads from immutable input arrays of $\mainProc$, making programs deterministic in our language~\cite{lahiri-cav12}. 
We desugar the program's global variables (including the heap) as additional input and output formal arguments to a procedure. 
We transform each procedure into its {\it Static Single Assignment} (SSA) form~\cite{ssa-cite}, where a variable is assigned at exactly one program node.


\subsection{Semantics}
\label{sec:semantics}


Let $\Values$ denote the set of values that variables and expressions can evaluate to.
Let $\sstore \in \Stores$ be a {\it store} mapping variables to values in $\Values$. 
For $\varn{x} \in \ProcVars{}$, we define $\varn{x} \in \sstore$ if $\varn{x}$ is a variable in the domain of $\sstore$.
For $\varn{x} \in \sstore$, $\sstoreAt{\varn{x}}$ denotes the value of variable $\varn{x}$.
The store $\mkStore{\varn{x} \rightarrow \val{}}$ represents a singleton store that maps $\varn{x}$ to  $\val{}$.
The store $\projStore{\store{}}{\ProcVars{1}}$ restricts the domain of the store to variables in $\ProcVars{1}$.
For stores $\store{1}$ and $\store{2}$, the store $\store{3} \doteq \storeUnion{\store{1}}{\store{2}}$ is defined as follows for any variable $\varn{x} \in \store{1}$ or $\varn{x} \in \store{2}$:
\[
\storeAt{3}{\varn{x}} = 
\begin{cases}
  \storeAt{2}{\varn{x}}, & \text{if } \varn{x} \in \store{2}\\
  \storeAt{1}{\varn{x}}, & \text{otherwise}
\end{cases}
\]
The value of an expression $e \in \Exprs$ ($\sstoreAt{e}$) is defined inductively on the structure of $e$ (we omit it for brevity as it is fairly standard).

\myparabf{Calls}~Let $\ccstck \in (\ProcNodes{} \times \Vars{}^* \times \Stores)^*$ be a {\it call stack} that is a sequence of tuples $\langle (\procNode{0}, \vect{\varn{r}}_0, \store{0}), (\procNode{1}, \vect{\varn{r}}_1, \store{1}), \ldots \rangle$, where $\procNode{i}$ is the $i$-th call site on the call stack ($\procNode{0}$ is the most recent), $\vect{\varn{r}}_i$ and  $\store{i}$, respectively, are the vector of return actuals and  the valuation of the local variables of the caller, at the corresponding call site.
Let $\Cstcks{}$ denote the set of all such call stacks, $\emptyStck$ denotes an empty stack, and $\concatStck{(\pprocNode, \vect{\varn{r}}, \sstore)}{\ccstck}$ denotes the {\it concatenation} operator.

\myparabf{Transition Relation}~A {\it state} $\sstate \in \States$ is a tuple $(\pprocNode, \sstore, \ccstck) \in \ProcNodes{} \times \Stores \times \Cstcks{}$ that denotes a point in program execution where $\pprocNode$ is the current node being executed in a procedure $\pproc$, $\sstore$ is the valuation of variables in $\ProcVars{\pproc}$ and $\ccstck$ is the current call stack.


%
%

A {\it state transition} denoted as $\step{(\procNode{\pproc}, \store{1}, \cstck{1})}{(\procNode{2}, \store{2}, \cstck{2})}$ is a relation over $\States \times \States$ holds only if:
\begin{enumerate}
\setlength\itemsep{\spcing}
\item $\stmtMap(\procNode{\pproc}) \doteq \bassign{\varn{x}}{e}$, 
$\procNode{2} \in \ProcNodes{\pproc}$, $\store{2} = \storeUnion{\store{1}}{\mkStore{\varn{x} \rightarrow \storeAt{1}{e}}}$, $(\procNode{\pproc}, \procNode{2}) \in \ProcEdges{\pproc}$, and $\cstck{1} = \cstck{2}$, or
\item $\stmtMap(\procNode{\pproc}) \doteq \bassume{e}$, 
$\procNode{2} \in \ProcNodes{\pproc}$, $\storeAt{1}{e} = \btrue$, $(\procNode{\pproc}, \procNode{2}) \in \ProcEdges{\pproc}$, $\store{1} = \store{2}$ and $\cstck{1} = \cstck{2}$, or
\item $\stmtMap(\procNode{\pproc}) \doteq \bskip$, $\procNode{\pproc} \neq \procExitNode{\pproc}$, $\procNode{2} \in \ProcNodes{\pproc}$, $(\procNode{\pproc}, \procNode{2}) \in \ProcEdges{\pproc}$, $\store{1} = \store{2}$ and $\cstck{1} = \cstck{2}$, or
\item $\stmtMap(\procNode{\pproc}) \doteq \bcall{\aaltproc}{\vect{e}}{\vect{\varn{r}}}$. 
Let $\pprocNode$ be the unique successor of $\procNode{\pproc}$ in $\pproc$, and $\vect{\varn{x}}$ be the vector of input formals for $\aaltproc$ in
$\procNode{2} = \procEntryNode{\aaltproc}$, $\cstck{2} = \concatStck{(\pprocNode, \vect{\varn{r}}, \store{1})}{\cstck{1}}$ and $\store{2} = \mkStore{\vect{\varn{x}} \rightarrow \storeAt{1}{\vect{e}}}$, or
\item $\stmtMap(\procNode{\pproc}) \doteq \bskip$, $\procNode{\pproc} = \procExitNode{\pproc}$, $\cstck{1} \doteq (\procNode{\aaltproc}, \vect{\varn{r}}, \store{3})::\cstck{3}$. 
Let $\vect{\varn{y}}$ be the vector of output formals for $\pproc$ in
$\procNode{2} = \procNode{\aaltproc}$, $\store{2} = \projStore{(\storeUnion{\store{3}}{\mkStore{\vect{r} \rightarrow \storeAt{1}{\vect{\varn{y}}}}})}{\ProcVars{\aaltproc}}$, $\cstck{2} = \cstck{3}$. 
\end{enumerate}

A transitive edge $\stepTransitive{\state{0}}{\state{n}}$ exists if $\state{n} \equiv \state{0}$ or there exists a sequence of transitions $\step{\state{0}}{\state{1}}, \ldots \step{\state{n-1}}{\state{n}}$, where $\step{\state{i}}{\state{i+1}}$, for all $i \in [0,\ldots,n)$.
For a procedure $\pproc$, we denote the input-output {\it transition relation} $\transRel{\pproc} \doteq \{(\store{1}, \store{2}) \ | \ \stepTransitive{(\procEntryNode{\pproc}, \store{1}, \emptyStck)}{(\procExitNode{\pproc}, \store{2}, \emptyStck)}\}$.


\myparabf{Execution Traces}~An {\it execution trace} $\trace{}$ is a (possibly infinite) sequence of states $\langle \state{0}, \state{1}, \ldots \rangle$, where $\step{\state{i}}{\state{i+1}}$, for any adjacent pair of states in the sequence.
For a trace $\trace{}$ and a node $\pprocNode \in \ProcNodes{}$, $\projectTraceOnNode{\trace{}}{\pprocNode}$ denotes the (maximal) subsequence of $\trace{}$ containing states of the form $(\pprocNode, \_, \_)$.
%
For such a trace $\trace{}$ of length at least $i+1$,  $\traceAt{}{i}$ denotes the state at position $i$ (namely $\state{i}$).
For any procedure $\pproc$, let $\ProcTraces{\pproc}$ be the set of {\it maximal} traces of $\pproc$.
That is, $\ProcTraces{\pproc}$ is the set of all traces $\trace{}$ such that
(i) $\traceAt{}{0} \doteq (\procEntryNode{\pproc}, \_, \emptyStck)$, and 
(ii) either (a) the final state $\state{n}$ has no successors, or (b) the trace is non-terminating.
Traces with no successors can either terminate {\it normally} in a state $(\procExitNode{\pproc}, \_, \emptyStck)$, or could be {\it blocked} due to no successors in $\ProcEdges{}$ or due to an unsatisfied $\bassumen$ statement. 
For a store $\sstore \in \Stores{}$, we denote $\ProcTracesStarting{\pproc}{\sstore}{}$ as the maximal trace (due to determinism) of $\pproc$ that starts in a store $\sstore$.

%% file: change.tex

\section{Problem Statement}
\label{sec:problem}

In this section, we formalize the problem of {\it semantic change-impact analysis} and provide a simple solution based on dataflow-based static analysis. 

%

\subsection{Representing Changes}
\label{sec:change}
%

We denote $\program{1}, \program{2} \in \Programs$ as two versions of a program.
Similarly $\sstate^i, \sstore^i, \trace{}^i, \Procs^i, \mainProc^i, \stmtMap^i$ denote entities for $\program{i}$, without making $\program{i}$ explicit.

To simplify the formulation we assume the two programs in a {\it normalized} form, where 
(i) each procedure in $\Procs^1$ has a corresponding procedure in $\Procs^2$ and vice versa, and
(ii) for each  $\pproc \in \Procs^i$, the vector of variables in $\ProcVars{\pproc}$, and the set of nodes $\ProcNodes{\pproc}$ (but not necessarily $\ProcEdges{\pproc}$) are identical with the ones in the corresponding procedure.
We can easily preprocess the programs to obtain their normalized form, by introducing additional procedures, variables (uninitialized) and nodes.
Finally, for any missing node $\procNode{}$, we add an unreachable node in $\ProcNodes{\pproc}$ with a $\bskip$ statement and empty successor list.


\myparabf{Diffing}~Given the two versions, a diffing algorithm produces a mapping between nodes in the two programs. We assume we are given a sound diff algorithm to label the sources of change.
A diff algorithm is sound if it produces a partial function $\map : \ProcNodes{1} \pfun \ProcNodes{2}$ such that:
\begin{enumerate}
\setlength\itemsep{\spcing}
\item $\map$ is a {\emph partial} bijection\footnote{A partial bijection is a partial function that is injective when defined and (trivially) surjective when restricted to its image~\cite{grillet1995semigroups}.} and $\stmtMap(\procNode{\pproc}) =
 \stmtMap(\map(\procNode{\pproc}))$.
\item For any two traces $\trace{}^1 \doteq \ProcTracesStarting{\mainProc}{\sstore}{1}$ in $\program{1}$ and $\trace{}^2 \doteq \ProcTracesStarting{\mainProc}{\sstore}{2}$ in $\program{2}$, 
$\trace{}^1$ only executes statements in $Dom(\map)$ 
iff $\trace{}^2$ only executes statements in $Im(\map)$ 
\item 
For any two traces $\trace{}^1 \doteq \ProcTracesStarting{\mainProc}{\sstore}{1}$ in $\program{1}$ and $\trace{}^2 \doteq \ProcTracesStarting{\mainProc}{\sstore}{2}$ in $\program{2}$,
where $\trace{}^1$ only executes statements in $Dom(\map)$ 
or $\trace{}^2$ only executes statements in $Im(\map)$,
then $\trace{}^1 = \trace{}^2$.
\end{enumerate}

\newcommand{\changedProcs}{\Procs{}^\Delta}

The mapped nodes $\mappedNodes{} \doteq Dom(\map) \cup Im(\map)$ underapproximate the set of nodes that are syntactically unchanged.
Intuitively, if a program executes only statements in $\mappedNodes{}$ then the program behaves the same in both versions;
statements that are not in $\mappedNodes{}$ are the sources of change.

We describe for illustrative purposes a simple diffing algorithm which is sound.
The algorithm proceeds to produce a mapping $\map$ as follows:
Let $\changedProcs \subseteq \Procs{}$ be the set of procedures that have some syntactic change. 
Any node not in $\pproc \in\changedProcs$ is trivially mapped as the control-flow graphs are identical in the two versions. 
Any node in $\pproc \in \changedProcs{}$ is conservatively treated as not mapped.
Our formulation is parameterized by a diff algorithm which can either be based on text~\cite{yang1991identifying} or more  sophisticated notions such as abstract syntax trees~\cite{falleri2014fine} or program-dependency-graphs~\cite{krinke2001identifying} as long as they satisfy the soundness criteria.




\subsection{Semantic Change Impact}
\label{sec:semantic-impact}

We can now state the meaning of a node being impacted by a program change, in terms of the trace semantics of the two programs and the set $\mappedNodes$.

For a sequence of states $\vect{\sstate}$ and a variable $\varn{x} \in \ProcVars{}$, $\varSeqInStateSeq{\vect{\sstate}}{\varn{x}} \in (\Values \cup \{\botValue\})^*$ denotes the sequence of values $\vect{\val{}}$ with the same length as $\vect{\sstate}$, and 
\vspace{-0.3em}
\[
\vspace{-0.5em}
\val{i} = 
\begin{cases}
   \sstoreAt{\varn{x}}, & \state{i} \doteq (\_, \sstore, \_)  \text{ and } \varn{x} \in \sstore \\
   \botValue & \text{otherwise}
\end{cases}
\]

%
%

\begin{definition}[Impacted nodes]
Given $\program{1}, \program{2}$ and $\mappedNodes{}$, a node $\pprocNode \in \ProcNodes{}^{1}\cup\ProcNodes{}^{2}$ is {\it impacted} if either
$\impacted{\pprocNode}{\program{1}}{\program{2}}{\map}$ or
$\impacted{\map(\pprocNode)}{\program{2}}{\program{1}}{\invMap}$ holds, where $\invMap$ is the inverse.
$\ProcNodes{}^{i}$ is the corresponding $\ProcNodes{}$ for $\program{i}$.

We define $\impacted{k}{\program{a}}{\program{b}}{\vmap}$:
\begin{enumerate}
\setlength\itemsep{\spcing}
\item $k \not\in Dom(\vmap)$, or
\item there exists a store $\sstore$, pair of traces $\trace{}^a \doteq \ProcTracesStarting{\mainProc}{\sstore}{a}$ for $\program{a}$ and $\trace{}^b \doteq \ProcTracesStarting{\mainProc}{\sstore}{b}$ for $\program{b}$, and a variable $\varn{x} \in \readSet{\pprocNode}$ such that $\varSeqInStateSeq{(\projectTraceOnNode{\trace{}^a}{k})}{\varn{x}} \neq \varSeqInStateSeq{(\projectTraceOnNode{\trace{}^b}{\vmap(k)})}{\varn{x}}$.
\end{enumerate}
\label{defn:impacted}
\end{definition}

\begin{table}
\begin{center}
\scriptsize
\begin{tabular}{|l|l|}
\hline
\textbf{Predicate name} & \textbf{Definition} \\ \hline \hline 
$\conditionalNode{\pprocNode}$ &  if $\pprocNode$  is a branching node \\ \hline
$\controlDep{\procNode{1}}{\procNode{2}}$ &  if  $\procNode{2}$ is {\it control-dependent} on $\procNode{1}$~\cite{ferrante-toplas87} \\ \hline
$\callsite{\pprocNode}{\pproc}{\aaltproc}$ &  if $\stmtMap(\pprocNode)$ is a call to $\pproc$ within\\
& a caller $\aaltproc$. \\ \hline
$\inParam{\varn{x}}{i}{\pproc}$ &  if $\varn{x}$ is the $i$-th input formal of $\pproc$ \\ \hline
$\outParam{\varn{x}}{i}{\pproc}$ &  if $\varn{x}$ is the $i$-th output formal of $\pproc$ \\ \hline
$\inActual{e}{i}{\pproc}{\pprocNode}$ &  if the expression $e$ is the $i$-th \\
& actual argument   to a call to $\pproc$\\
& at a callsite $\pprocNode$ \\ \hline
$\outActual{\varn{r}}{i}{\pproc}{\pprocNode}$ &  if the variable $\varn{r}$ receives the $i$-th  \\
& output formal  to a call to $\pproc$\\
& at a callsite $\pprocNode$ \\ \hline
\end{tabular}
\end{center}
\caption{Predicates used for dataflow analysis.}
\label{tab:predicates-dataflow}
\vspace{\vsp}
\vspace{\vsp}
\vspace{\vsp}
\end{table}

We conservatively treat any unmapped node as impacted.
A mapped node $\procNode{}$ is not impacted if the sequence of values of variables in $\readSet{\pprocNode}$ is identical for any two execution traces $\trace{}^a$ (in $\program{a}$) and $\trace{}^b$ (in $\program{b}$) starting from a common input store $\sstore$ to $\mainProc$.
Note that for our low-level language, the $\readSet{\pprocNode}$ of a statement often includes the state of the heap and address being written to.
For example, the C\# statement $\varn{x.length} = \varn{y}$ gets translated to $\pprocNode: \bassign{\varn{Length}}{\updateFunc{\varn{Length}}{\varn{x}}{\varn{y}}}$, (where $\varn{Length}$ is an array variable representing the state of $\varn{length}$ field/attribute in all objects) with $\readSet{\pprocNode} = \{\varn{Length}, \varn{x}, \varn{y}\}$.




\subsection{Dataflow-based Change-Impact Analysis}
\label{sec:dataflow}

\SetKwFunction{dataflowcia}{DCIA}
\SetKwFunction{semDataflowcia}{SEM-DCIA}
\SetKwFunction{anytimeAlgo}{SEM-DCIA-ANYTIME}
\SetKwFunction{InferEq}{InferEquivs}

In this section, we describe {\it Dataflow-based Change-Impact Analysis} ($\dataflowcia$), a {\it change semantics unaware} static analysis that provides a conservative estimate of the set of impacted nodes.
The static analysis is an interprocedural dataflow analysis~\cite{ipdfs-cite} that starts with a program $\program{i}$ ($i \in {1,2}$) and a conservative estimate of the syntactically-changed nodes, nodes not in $\mappedNodes$, and returns an upper bound on the set of  (a) impacted nodes, (b) impacted variables, and (c) output variables whose summary may have changed.

\Comment{We assume that each procedure is transformed into a {\it Static Single Assignment} (SSA) form~\cite{ssa-cite}, where a variable (instance) is assigned at exactly one program node.
We extend $\ProcVars{}$ to include the SSA-renamed instances of existing variables.
Due to the SSA form and the lack of global variables, a variable $\varn{x} \in \ProcVars{}$ uniquely identifies the enclosing procedure and the node where the variable is written to.}

%
%

\myparabf{Predicates}
Table~\ref{tab:predicates-dataflow} defines some straightforward predicates used in the inference rules.
The $\outActual{\varn{r}}{i}{\pproc}{\pprocNode}$ predicate holds when the $i^{th}$ return value is assigned to variable $r$, at the call to $\pproc$ from the node $\pprocNode$ (note that we allow multiple return values); we call $r$ the output actual to differentiate it from the $i^{th}$ output formal inside the callee.
For $\controlDep{\procNode{1}}{\procNode{2}}$, a node $\procNode{2}$ is control-dependent on node $\procNode{1}$ iff  (i)  there exists a path from $\procNode{1}$ to $\procNode{2}$ s.t.  every  node in the path other than $\procNode{1}$ and $\procNode{2}$ is {\it post-dominated} by $\procNode{2}$, and  (ii) $\procNode{1}$ is not post-dominated by $\procNode{2}$~\cite{ferrante-toplas87}. 

\newcommand{\InfRuleDependsEntry}{\textsc{DEPENDS-ENTRY}}
\newcommand{\InfRuleDependsWrite}{\textsc{DEPENDS-WRITE}}
\newcommand{\InfRuleDependsTransitive}{\textsc{DEPENDS-TRANSITIVE}}
\newcommand{\InfRuleDependsNode}{\textsc{DEPENDS-NODE}}
\newcommand{\InfRuleDependsControl}{\textsc{CONTROL-DEPENDS}}
\newcommand{\InfRuleDependsSummary}{\textsc{SUMMARY-DEPENDS}}

\begin{figure}[htbp]
{\small{
\begin{mathpar}
\inferrule[\InfRuleDependsEntry]
{\varn{x} \in \ProcIns{\pproc}}
{\dependsOnVar{\varn{x}}{\varn{x}}{\pproc}}

\inferrule[\InfRuleDependsWrite]
{\varn{x} \in \readSet{\pprocNode} \\ \varn{y} \in \writeSet{\pprocNode} \\ \pprocNode \in \ProcNodes{\pproc}}
{\dependsOnVar{\varn{y}}{\varn{x}}{\pproc}}

\inferrule[\InfRuleDependsTransitive]
{\dependsOnVar{\varn{y}}{\varn{x}}{\pproc} \\ \dependsOnVar{\varn{x}}{\varn{z}}{\pproc}}
{\dependsOnVar{\varn{y}}{\varn{z}}{\pproc}}

\inferrule [\InfRuleDependsControl]
{\conditionalNode{\procNode{1}} \\ \varn{x} \in \readSet{\procNode{1}} \\ \controlDep{\procNode{1}}{\procNode{2}} \\ \varn{y} \in \writeSet{\procNode{2}}}
{\dependsOnVar{\varn{y}}{\varn{x}}{\pproc}}

\inferrule [\InfRuleDependsSummary]
{\callsite{\pprocNode}{\pproc}{\aaltproc} \\ \outActual{\varn{r}}{i}{\pproc}{\pprocNode} \\ \outParam{\varn{y}}{i}{\pproc} \\ \inParam{\varn{x}}{j}{\pproc} \\ \dependsOnVar{\varn{y}}{\varn{x}}{\pproc} \\  \inActual{e}{j}{\pproc}{\pprocNode} \\ \varn{w} \in \readSet{e}}
{\dependsOnVar{\varn{r}}{\varn{w}}{\aaltproc}}

\inferrule[\InfRuleDependsNode]
{\varn{x} \in \writeSet{\pprocNode} \\ \pprocNode \in \ProcNodes{\pproc} \\ \dependsOnVar{\varn{y}}{\varn{x}}{\pproc}}
{\dependsOnNode{\varn{y}}{\pprocNode}{\pproc}}

\end{mathpar}
}}
\vspace{\vsp}
\caption{Inference rules for computing $\dependsOnVarn$ and $\dependsOnNoden$. The input is a program $\program{}$.}
\label{fig:depends}
\vspace{\vsp}
\vspace{\vsp}
\end{figure}

\myparabf{Dependency}
Figure~\ref{fig:depends} describes a set of inference rules to compute two relations $\dependsOnVarn$ and $\dependsOnNoden$.
For a pair of variables $\varn{x}, \varn{y} \in \ProcVars{\pproc}$ such that $\varn{y}$ is either data- or control-dependent on $\varn{x}$, then $\dependsOnVar{\varn{y}}{\varn{x}}{\pproc}$ holds.
Similarly, a node $\pprocNode \in \ProcNodes{\pproc}$ and a variable $\varn{y}$ such that $\varn{y}$ is data or control dependent on a variable $\varn{x}$ that is updated at $\pprocNode$, $\dependsOnNode{\varn{y}}{\pprocNode}{\pproc}$ holds. 
%
An inference rule (e.g. \InfRuleDependsNode) lists a set of antecedents (above the line) and the consequent (below the line). 
Applying an inference rule results in adding a tuple to the relation in the consequent (e.g. $\dependsOnNoden$)\Comment{ to satisfy the consequent}.
The inference rules are applied repeatedly until a fix-point is reached.

Most of the inference rules are straightforward encoding of program data- and control flow.
The rule \InfRuleDependsControl{} expresses that if $\procNode{1}$ is a branching node, whose condition depends on $\varn{x}$ and $\varn{y}$ is written in a control-dependent node $\procNode{2}$, then $\varn{y}$ depends on $\varn{x}$. 
The rule \InfRuleDependsSummary{} captures the dependency of an actual return $\varn{r}$ on a variable $\varn{w}$ passed as an argument to $\pproc$ in a caller $\aaltproc$, where $\varn{w}$ indirectly flows to $\varn{r}$ through a procedure call to $\pproc$.
For this callsite, the $i$-th output formal $\varn{y}$  (which is assigned to the output actual $\varn{r}$) is dependent on the $j$-th input formal $\varn{x}$, which in turn is assigned the actual $e$ at the callsite.

\newcommand{\TaintInfRuleSynt}{\textsc{SYNT-CHANGED}}
\newcommand{\TaintInfRuleNodeToVar}{\textsc{NODE-2-VAR}}
\newcommand{\TaintInfRuleVarToExpr}{\textsc{VAR-2-EXPR}}
\newcommand{\TaintInfRuleVarToNode}{\textsc{VAR-2-NODE}}
\newcommand{\TaintInfRuleTaintSumm}{\textsc{IMPACT-SUMMARY}}
\newcommand{\TaintInfRuleTaintSummProp}{\textsc{IMPACT-SUMMARY-PROP}}
\newcommand{\TaintInfRuleCallTaint}{\textsc{CALL-IMPACT}}
\newcommand{\TaintInfRuleReturnTaint}{\textsc{RETURN-IMPACT}}
\newcommand{\TaintInfRuleSummTaint}{\textsc{SUMMARY-IMPACT}}

\begin{figure*}[htbp]
{\small{
\begin{mathpar}
\inferrule [\TaintInfRuleSynt]
{\notmapped{\pprocNode}}
{\taintedNode{\pprocNode}}

%
\inferrule [\TaintInfRuleNodeToVar]
{\taintedNode{\pprocNode} \\ \varn{x} \in \writeSet{\pprocNode}}
{\taintedVar{\varn{x}}}

%
\inferrule [\TaintInfRuleVarToExpr]
{\taintedVar{\varn{x}} \\ \varn{x} \in \readSet{e}}
{\taintedExpr{e}}

%
\inferrule [\TaintInfRuleVarToNode]
{\taintedVar{\varn{x}} \\ \varn{x} \in \readSet{\pprocNode}}
{\taintedNode{\pprocNode}}

\inferrule[\TaintInfRuleTaintSumm]
{\outParam{\varn{y}}{i}{\pproc} \\ \dependsOnNode{\varn{y}}{\pprocNode{}}{\pproc} \\ \pprocNode{} \not\in \mappedNodes{} \\ \highlight{\neg \summEq{\varn{y}}{\pproc}}}
{\taintedSumm{\varn{y}}{\pproc}}

\inferrule[\TaintInfRuleTaintSummProp]
{\outParam{\varn{y}}{i}{\pproc} \\ \callsite{\pprocNode}{\aaltproc}{\pproc} \\ \outParam{\varn{x}}{j}{\aaltproc} \\ \taintedSumm{\varn{x}}{\aaltproc} \\ \outActual{\varn{w}}{j}{\aaltproc}{\pprocNode} \\ \dependsOnVar{\varn{y}}{\varn{w}}{\pproc} \\ \highlight{\neg \summEq{\varn{y}}{\pproc}}}
{\taintedSumm{\varn{y}}{\pproc}}

\inferrule [\TaintInfRuleCallTaint]
{\callsite{\pprocNode}{\pproc}{\aaltproc} \\ \inActual{e}{i}{\pproc}{\pprocNode} \\ \taintedExpr{e} \\ {\inParam{\varn{x}}{i}{\pproc}} \\ \highlight{\neg \preEq{\varn{x}}{\pproc}}}
{\taintedVar{\varn{x}}}

\inferrule [\TaintInfRuleReturnTaint]
{\callsite{\pprocNode}{\pproc}{\aaltproc} \\ \outActual{\varn{r}}{i}{\pproc}{\pprocNode} \\ \outParam{\varn{y}}{i}{\pproc} \\ \taintedSumm{\varn{y}}{\pproc} } 
{\taintedVar{\varn{r}}}

\inferrule [\TaintInfRuleSummTaint]
{\callsite{\pprocNode}{\pproc}{\aaltproc} \\ \outActual{\varn{r}}{i}{\pproc}{\pprocNode} \\ \outParam{\varn{y}}{i}{\pproc} \\ \inParam{\varn{x}}{j}{\pproc} \\ \dependsOnVar{\varn{y}}{\varn{x}}{\pproc} \\  \inActual{e}{j}{\pproc}{\pprocNode} \\ \taintedExpr{e} \\ \highlight{\neg (\preEq{\varn{x}}{\pproc} \wedge \summEq{\varn{y}}{\pproc})}}
{\taintedVar{\varn{r}}}
\end{mathpar}
}} 
\vspace{\vsp}
\caption{Inference rules for dataflow based change-impact analysis.
The \highlight{\text{highlighted}} antecedents are relevant for change-semantics aware analysis.
}
\label{fig:dataflow}
\vspace{\vsp}
\end{figure*}
    
\myparabf{Impact Analysis}
Figure~\ref{fig:dataflow} describes a set of inference rules to compute the set of nodes that are impacted in either program.
For this section, we will ignore the \highlight{\text{highlighted}} antecedents (they become relevant in \S~\ref{sec:equiv} where we describe how we incorporate change semantics).
The rules take as input a program (either $\program{1}$ or $\program{2}$), the set of mapped nodes $\mappedNodes$ and precomputed relations $\dependsOnNoden$ and $\dependsOnVarn$ for the particular program.
They produce the relations \taintedNoden{}, \taintedVarn{} and \taintedSummn{} that represent an upper bound on the set of impacted nodes, impacted variables and impacted variable summaries, respectively.
Most rules are self-explanatory; we describe the main rules relevant to the interprocedural reasoning. 
Note that we do not have a special rule for control-flow impact, since $\dependsOnVarn$ already captures control flow dependency.

For an output formal $\varn{y} \in \ProcOuts{\pproc}$, the summary (input-output dependency) may change either when (i) $\varn{y}$ depends on a variable updated at an unmapped node $\pprocNode \in \ProcNodes{\pproc}$ (expressed by \TaintInfRuleTaintSumm), or (ii) $\varn{y}$ depends on a variable $\varn{w}$ that stores the output formal $\varn{x}$ of $\aaltproc$ at a callsite in node $\pprocNode$, and the summary of $\varn{x}$ has changed in $\aaltproc$ (expressed by \TaintInfRuleTaintSummProp).

The \TaintInfRuleCallTaint{} rule says that an input formal $\varn{x}$ in $\pproc$ can be impacted if the corresponding actual argument $e$ at a callsite is impacted.
\TaintInfRuleReturnTaint{} considers the case when  the variable summary for the corresponding output formal $\varn{y}$ is impacted.
\TaintInfRuleSummTaint{} considers the case when the actual argument expression $e$ (passed for the formal $\varn{x}$ of $\pproc$) is impacted, and $\varn{y}$ depends on the value of $\varn{x}$ in $\pproc$.
Our analysis preserves context-sensitivity as it does not impact a return value simply because the corresponding output formal is impacted in some context.


The algorithm $\dataflowcia$ does the following:
\begin{enumerate}
\setlength\itemsep{\spcing}
\item Takes as input $\program{1}, \program{2}$ and $\mappedNodes$.
\item Applies the inference rules in Figure~\ref{fig:dataflow} on $\program{i}$ to generate $\taintedNoden^i$, $\taintedVarn^i$, $\taintedSummn^i$ until a fix-point is reached.
\item Returns the tuple ($\bigcup_i \taintedNoden^i$, $\bigcup_i \taintedVarn^i$,  $\bigcup_i \taintedSummn^i$). 
\end{enumerate}

The following theorem states the soundness of the dataflow analysis $\dataflowcia$. 
\begin{theorem}[Soundness]
Given $\program{1}, \program{2} \in \Programs$ and $\mappedNodes \subseteq \ProcNodes{}$, (a) $\dataflowcia$ terminates, and 
(b) for any $\pprocNode \not\in \taintedNoden$, $\pprocNode$ is not an impacted node with respect to \mappedNodes~(according to Definition~\ref{defn:impacted}).
\label{thm:dcia}
\end{theorem}

Consider the changes in \fig~\ref{fig:running-example} at line~\ref{line:change30};
the procedure \codein{locale\_ok} has an impacted summary because its return variable depends on a node that is syntactically changed, i.e., is not in $\mappedNodes$. 
This causes the line~\ref{line:localeok} and the variable  \codein{locale} to be marked as impacted because of the rule $\TaintInfRuleTaintSumm$. Impacts are propagated interprocedurally by the rule $\TaintInfRuleCallTaint$ to all calls that take \codein{locale} as an argument, i.e., \codein{print\_name}, \codein{print\_major\_version}, and \codein{print\_minor\_version}.
Similarly, using the same rule, the body of \codein{print\_header} is impacted by the changed argument \codein{`\textbackslash{}n'} changed to the variable \codein{line\_delim} on line~\ref{line:change11}.
The propagation through calls further impacts their entire body because of the data and control dependency on the impacted argument (by the rules $\TaintInfRuleNodeToVar$ and $\TaintInfRuleVarToNode$ which propagate impact through both control- and data-dependency relying on the predicate $\dependsOnVarn$).  

%
%


%% file: technique.tex

\section{Incorporating Change Semantics}
\label{sec:equiv}

In this section, we make the $\dataflowcia$ algorithm {\it change-semantics aware}. 
In other words, the analysis takes into account also the exact semantics of the change, in addition to the set of nodes $\mappedNodes$ that may have been syntactically changed.
We inject the change-semantics by leveraging equivalence relationships between variables and procedure summaries in the two programs $\program{1}$ and $\program{2}$.

%
%

Let us define the following semantic equivalences for a variable over $\program{1}$ and $\program{2}$.

\begin{definition}[$\preEqn$]
$\preEq{\varn{x}}{\pproc}$ holds for an input formal $\varn{x} \in \ProcIns{\pproc}$ if for all stores $\sstore$, and for every pair of traces $\trace{}^1 \doteq \ProcTracesStarting{\mainProc}{\sstore}{1}$ and $\trace{}^2 \doteq \ProcTracesStarting{\mainProc}{\sstore}{2}$, $\varSeqInStateSeq{(\projectTraceOnNode{\trace{}^1}{\procEntryNode{\pproc}})}{\varn{x}} = \varSeqInStateSeq{(\projectTraceOnNode{\trace{}^2}{\map(\procEntryNode{\pproc})})}{\varn{x}}$.
\label{defn:preEq}
\end{definition}

Intuitively, $\preEq{\varn{x}}{\pproc}$ holds for an input formal $\varn{x}$ of $\pproc$ if any two executions starting from the same input $\sstore$ to $\mainProc$ call $\pproc$  with the same sequence of values of $\varn{x}$.
For example, in \fig~\ref{fig:running-example} the equivalences $\preEq{\varn{delim}}{print\_header}$, $\preEq{\varn{locale}}{print\_name}$,  $\preEq{\varn{locale}}{print\_major\_version}$,  and $\preEq{\varn{locale}}{print\_minor\_version}$ hold. 
In contrast,  $\preEq{\varn{delim}}{print\_minor\_version}$ does \emph{not} hold, because of different values for \varn{delim} \codein{`\textbackslash{}n`} and \codein{`\textbackslash{}0`} respectively, at the call-site in ${print\_product\_info}$,.



Let us define $\dependencyOf{\varn{y}}$ as the set of variables $\varn{x}$ such that $\dependsOnVar{\varn{y}}{\varn{x}}{\pproc}$ in either $\program{1}$ or $\program{2}$. 
For two stores $\store{1}$ and $\store{2}$ defined over same set of variables, we denote $\equalOn{\store{1}}{\store{2}}{\ProcVars{1}}$ to mean $\storeAt{1}{\varn{x}} = \storeAt{2}{\varn{x}}$ for every $\varn{x} \in \ProcVars{1}$.

\begin{definition}[$\summEqn$]
$\summEq{\varn{y}}{\pproc}$ holds for an output formal $\varn{y} \in \ProcOuts{\pproc}$ if 
$(\store{1}, \store{2}) \in \transRel{\pproc}$ in $\program{i}$ and $\equalOn{\store{1}}{\store{3}}{\dependencyOf{\varn{y}}}$, then $(\store{3}, \store{4}) \in \transRel{\pproc}$ is in $\program{j}$ ($j \neq i$) and $\storeAt{2}{\varn{y}} = \storeAt{4}{\varn{y}}$.
\label{defn:summEq}
\end{definition}


Intuitively, if the versions of $\pproc$ are executed from stores $\store{1}$ and $\store{3}$ where $\equalOn{\store{1}}{\store{3}}{\dependencyOf{\varn{y}}}$, then either both procedures do not terminate, or the value of $\varn{y}$ after executing $\pproc$ is identical on exit.  In our example in \fig~\ref{fig:running-example}, all procedures are equivalent except \codein{print\_product\_info}, i.e., in this case $\summEq{\varn{line\_delim}}{print\_product\_info}$ does not hold since in one version the value of $line\_delim$ at the end of the execution is \codein{``\textbackslash 0''} while in the other it is undefined.

Figure~\ref{fig:dataflow} with the highlighted parts provides a refinement to the dataflow analysis to incorporate change semantics. 
In addition to the \mappedNodes, the algorithm now takes as input pre-computed relations $\preEqn$ and $\summEqn$.
In this section, we assume an oracle that provides these relations; we provide one implementation later (\S~\ref{sec:impl}).
The highlighted facts strengthen the antecedent of a rule and prevent it from being applicable in some contexts.
For example, the strengthened \TaintInfRuleCallTaint{} prevents an input formal $\varn{x}$ from being impacted if $\preEq{\varn{x}}{\pproc}$ holds.
Similarly, the strengthened \TaintInfRuleTaintSumm{} prevents a summary for $\varn{y}$ from impact if we know that $\summEq{\varn{y}}{\pproc}$ holds.
The strengthened \TaintInfRuleSummTaint{} is now applicable only when either (i) the formal $\varn{x}$ does not satisfy $\preEqn$ or (ii) the summary for $\varn{y}$ does not satisfy $\summEqn$.

We denote the  new change-semantics aware algorithm as {\it Semantic Dataflow-based Changed Impact Analysis} ($\semDataflowcia$).

\begin{theorem}[Soundness]
Given $\program{1}, \program{2} \in \Programs$, $\mappedNodes$, $\preEqn$, and $\summEqn$, (i) $\semDataflowcia$ terminates, and (ii) for any $\pprocNode \not\in \taintedNoden$, $\pprocNode$ is not an impacted node with respect to \mappedNodes~(from Definition~\ref{defn:impacted}).
\end{theorem}

\subsection{Anytime Algorithm}
\label{sec:anytime}

The $\semDataflowcia$ algorithm assumes an oracle to compute the $\preEqn$ and $\summEqn$ relations.
Computing such equalities typically require constructing the product of the two programs $\program{1}$ and $\program{2}$ and inferring equivalence relations over the product program~\cite{lahiri-fse13}.
Such inference algorithms typically have high complexity and therefore it is wise to apply them prudently.
In this section, we make a simple observation that allows us to interleave $\semDataflowcia$ and inference of $\preEqn$ and $\summEqn$ in a single framework. 

To exploit the change semantics, it is often useful to apply equivalence relation inference only in the vicinity of actual syntactic changes.

\begin{figure}[h!]
\vspace{\vsp}
\begin{minipage}[t]{0.2\textwidth}
\begin{lstlisting}[style=C-github,mathescape=true,numbers=none]
void $\mainProc$(int x) 
{
@d@- (*@{\btHL[fill=gitdel]$\proc{1}$}@*)(x);@d@ 
@a@+ (*@{\btHL[fill=gitadd]$\proc{1}$}@*)(x@af@+0@af@);@a@
}
void $\proc{1}$(int x) 
{
  $\proc{2}$(x+1);
}
\end{lstlisting}
\end{minipage}
\begin{minipage}[t]{0.25\textwidth}
\begin{lstlisting}[style=C-github,mathescape=true,numbers=none]
void $\proc{2}$(int x) 
{
  $\proc{3}$(x+2);
}
...
void $\proc{n}$(int x) 
{
  ;
}
\end{lstlisting}
\end{minipage}
\vspace{\vsp}
\caption{Motivating example for anytime algorithm.}
\label{fig:example-anytime-1}
\vspace{\vsp}
\end{figure}

Consider the example in \fig~\ref{fig:example-anytime-1} to make the intuition clear.
Applying $\dataflowcia$ will result in impacting all the nodes in the program as follows.
The modified call node for $\proc{1}$  in $\mainProc$ is not in $\mappedNodes$, which impacts input formal $\varn{x}$ of $\proc{1}$.
This in turn impacts the call to $\proc{2}$ and so on.
On the other hand, we can observe that $\preEqn$ and $\summEqn$ hold for each of the procedures because the change does not propagate outside the changed statement. 

For \fig~\ref{fig:example-anytime-1} it suffices to infer the equivalences on $\mainProc$ while {\it abstracting} the rest of the procedures from the expensive equivalence analysis.
Considering $\proc{1}$ has all callsites inside $\mainProc$ and that it does not have an impacted summary by rule $\TaintInfRuleTaintSumm$ after $\dataflowcia$ suffices to determine that $\preEq{\varn{x}}{\proc{1}}$ holds.
This information can be fed to $\semDataflowcia$ which will prune the impact for the input parameter of $\proc{1}$ which will prune the remaining impacts when performing a pure dataflow analysis.
Thus, we obtain a precise change-impact analysis by applying the equivalence inference only on a small subset of the procedures in the program.
Similarly, in \fig~\ref{fig:running-example} it suffices to analyze only the syntactically changed procedures and abstract away the others to obtain the most precise result; this is not the case in general because to infer the $\preEqn$ we need all call sites to be in scope, not only the syntactically changed procedures.

%
%
%
%
%
%
%

\begin{algorithm}
 \KwIn{$\program{1}, \program{2} \in \Programs{}$}
 \KwIn{$\changedProcs \subseteq \Procs{}$}
 \KwIn{$\mappedNodes \subseteq \ProcNodes{}$}
 \KwOut{$\iNodes \subseteq \ProcNodes{}$}
 \Begin{
    $k \leftarrow 0$\;
    $\eqs \leftarrow (\emptyset, \emptyset)$\;
    $(\iNodes, \ivars, \isumms) \leftarrow$
    $~~~~~~~~~\semDataflowcia(\program{1}, \program{2}, \mappedNodes, \eqs)$\;\label{line:anytime-dataflow1}
    $\kChangedProcs \leftarrow \changedProcs{}$\;
    \While{$\kChangedProcs \subset \Procs{}$} { \label{line:anytime-loop}
	    $\preqs \leftarrow \{(\varn{x}, \pproc) \ | \ \varn{x} \in \ProcIns{\pproc} \text{ and } \varn{x} \not\in \ivars\}$\; \label{line:anytime-preqs}
            $\summeqs \leftarrow \{(\varn{x}, \pproc) \ | \ \varn{x} \in \ProcOuts{\pproc} \text{ and }$
            $~~~~~~~~~(\varn{x},\pproc) \not\in \isumms\}$\; \label{line:anytime-summeqs}
            $\eqs \leftarrow \eqs + (\preqs, \summeqs)$\; \label{line:anytime-unioneqs}
            $\kChangedProcs \leftarrow$
            $~~~~~~~~~\ProcsWithin(\changedProcs{}, \program{1}, \program{2}, k)$\; \label{line:anytime-procsk}
            $\program{1}_k \leftarrow \DropProcs(\program{1}, \Procs{} \setminus \kChangedProcs)$\; \label{line:anytime-dropprocs1}
            $\program{2}_k \leftarrow \DropProcs(\program{2}, \Procs{} \setminus \kChangedProcs)$\; \label{line:anytime-dropprocs2}
	    $\eqs \leftarrow \InferEq(\program{1}_k, \program{2}_k, \eqs)$\; \label{line:anytime-infereq}
	    $(\iNodes, \ivars, \isumms) \leftarrow$
	    $~~~~~~~~~\semDataflowcia(\program{1}, \program{2}, \mappedNodes, \eqs)$\; \label{line:anytime-dataflow2}
            $k$++ \;
    } 
    \Return \iNodes\;
 }
 \caption{SEM-DCIA-ANYTIME}
 \label{algo:anytime}
\end{algorithm}

Algorithm~\ref{algo:anytime}  (\anytimeAlgo) provides an {\it anytime} algorithm  that performs the integration.
The algorithm takes as an additional input $\changedProcs{}$, the set of syntactically changed procedures. 
It outputs a set of  nodes $\iNodes$ that overapproximates the set of impacted nodes.
We term the algorithm anytime~\cite{dean1988analysis,zilberstein1995approximate, anytime-wiki} because the algorithm can be stopped at any time after the first call to \semDataflowcia to obtain a conservative bound for the impacted nodes.

The algorithm starts with invoking \semDataflowcia on the two programs with an empty set of equivalences in $\eqs$ (line~\ref{line:anytime-dataflow1}); this is identical to calling $\dataflowcia$. 
The return values provide a conservative measure on impacted variables, nodes and summaries respectively (Theorem~\ref{thm:dcia}).
The algorithm implements a loop (line~\ref{line:anytime-loop}) where it increases the frontier of procedures $\kChangedProcs$ around $\changedProcs{}$ that are analyzed for inferring equivalences in \InferEq (line~\ref{line:anytime-infereq}). 
Lines~\ref{line:anytime-preqs} and ~\ref{line:anytime-summeqs} construct equivalences from the provably non-impacted variables and summaries. 
These equivalences are added to $\eqs$ in line~\ref{line:anytime-unioneqs}.
\ProcsWithin returns all procedures that can reach or be reached from $\changedProcs{}$ within a call stack of depth $k$; $k$ is incremented with each iteration of the loop.
\DropProcs abstracts all procedures outside $\kChangedProcs$; it only retains the knowledge of whether any procedure $\pproc \in \kChangedProcs$ has additional call sites outside $\kChangedProcs$ - this determines whether $\preEqn$ can be inferred for a procedure.
\InferEq is invoked with a set of equivalences in $\eqs$ on the smaller programs $\program{i}_k$.
The final call to \semDataflowcia is used to compute the more refined set of impacted variables, nodes and summaries based on the equivalences discovered from \InferEq.
The loop terminates when $\kChangedProcs$ consists of the entire program; at this point \InferEq has already looked at the entire program and no new equivalences will be discovered in line~\ref{line:anytime-infereq}.

Let us denote $\semDataflowcia_k$ as an instantiation of the algorithm \anytimeAlgo that terminated after the loop is executed exactly $k+1$ times. 
We also denote $\semDataflowcia_\infty$ if the loop terminates normally after $\kChangedProcs$ equals $\Procs{}$. 

\begin{theorem}[Soundness]
Given $\program{1}, \program{2} \in \Programs$, $\mappedNodes$, and $\changedProcs$, if $\semDataflowcia_k$ terminates then
for any $\pprocNode \not\in \iNodes$, $\pprocNode$ is not an impacted node with respect to \mappedNodes{}  (according to Definition~\ref{defn:impacted}).
\end{theorem}

%% file: evaluation.tex

\section{Implementation and Evaluation}
\subsection{Implementation}
\label{sec:impl}

We implemented our \semimp analysis for C programs, but our analysis works on the intermediate verification language Boogie~\cite{boogie}.
We leverage SMACK~\cite{smack} to convert LLVM bytecode to Boogie programs.

%
\myparabf{Diffing}
For our initial implementation, we leveraged $\texttt{diff}$ over C files to produce the source of changes, i.e., nodes not in $\mappedNodes$.
However, $\texttt{diff}$ does not satisfy the soundness criteria for diff (Section~\ref{sec:change}) because of changes in macros, data structures, control-flow changes, etc.; we therefore conservatively consider all nodes in a changed procedure as sources of impacts. 
Although this can overapproximate the initial source of impact, the use of equivalences in \semimp allows us to prune the spurious impacts from escaping the syntactically-changed procedures;
All our code and scripts are available in the \SymDiff{} repository at: \url{https://symdiff.codeplex.com/}.

\myparabf{Inference}
We used \SymDiff{} to construct a product program and infer valid $\preEqn$ and $\summEqn$.
Given $\program{1}$ and $\program{2}$, \SymDiff{} generates a product program $\prodProg{1}{2}$ that defines a procedure $\prodProc{\pproc}{1}{2}$ for every $\pproc$ and $\map(\pproc) \in \Procs^i$.
For the product program $\prodProg{1}{2}$, one can leverage any of the (single program) invariant generation techniques to infer preconditions, postconditions (including two-state postconditions) on $\prodProc{\pproc}{1}{2}$.
Such invariants are {\it relational} in that they are over the state of two programs $\program{1}$ and $\program{2}$, and include equivalences relations such as $\preEqn$ (preconditions of $\prodProc{\pproc}{1}{2}$) and $\summEqn$ (summary of $\prodProc{\pproc}{1}{2}$).
To ensure our inferred equivalences are valid we require the programs to be equi-terminating~\cite{hawblitzel-cade13}; this is an area of future work -- for now we assume that changes do not introduce non-termination.
We modified \SymDiff{} to add candidates for inferring summaries and take as input cheaply-inferred equalities from \dataflowcia. More details can be found in our extended report~\cite{extended-report}.



\subsection{Evaluation}
In this section we demonstrate the effectiveness of our approach on GitHub projects with real program changes and also standard benchmark programs with artificial changes.
We show that our semantic based analysis, $\semDataflowcia$ improves on $\dataflowcia$ by reducing the size of the impacted set.
The size of the impacted set is a proxy metric for the effort necessary to perform many software engineering tasks such as code review and testing.

\input{tables/projects_summary}

We analyze \numGitChanges{} actual changes consisting of refactorings, feature additions, buggy changes, and bug fixes from  \numGithub{} projects from the GitHub repository. 
The projects, number of versions used, their size in source lines of code (SLOC), and corresponding change sizes (in number of C source lines changed) are summarized in \tab~\ref{tab:proj_summary}.
Our subjects are applications written in C such as a virtual machine program (\Use{pn.tinyvm}), a histogram creator (\Use{pn.histo}), a markdown presentation tool (\Use{pn.mdp}), a file-descriptor management library (\Use{pn.flingfd}) and a test-generation library (\Use{pn.theft}). 
In addition, we also include \numBench{} standard benchmark programs widely used by prior research on regression testing~\cite{hutchins1994experiments}. 
These benchmarks consist of \numSiemChanges{} manually introduced changes representing non-trivial and hard to detect bugs.
Our projects are sized between 142 lines of source code and 6205 (SLOC). 
The changes in our projects vary in size between very small changes, consisting of single line changes and larger ones, consisting of over 400 lines (most of our changes are on the small end of this spectrum).

For our experiments, we first compare \semimp against \dataflowcia to study the impact of adding change-semantics to the impact analysis (\S~\ref{sec:expts-cia}).
Next, we evaluate the cost-precision tradeoff of the anytime algorithm \anytimeAlgo (\S~\ref{sec:eval:incremental}). 
Finally, we present several representative examples discovered while applying our tool (\S~\ref{sec:anecdotes}).


\input{tables/results_dac_linear}

\subsection{Change-Semantic Aware Analysis}
\label{sec:expts-cia}

\tab~\ref{tab:results} shows the results of running our \semimp analysis on each of our subjects.
For each change, we measure the number of lines reported as impacted by dataflow analysis (columns \dataflowcia Impact) and also by \semimp (columns $\semDataflowcia_{\infty}$).
The columns $\semDataflowcia_i$ denote various bounds for \anytimeAlgo  and results will be discussed in \S~\ref{sec:eval:incremental}.
We report for each project the minimum and maximum number of impacted lines (min, max), and for the \semimp analysis we report also the average reduction of the size of the impacted set. 
Note that \semimp analysis always reports a subset of the set reported by the non-semantic analysis. 
We also report the average analysis time in seconds for the non-semantic analysis and for the \semimp analysis.

Our evaluation shows that on average, the change-aware analysis reduces the size of the impacted set by \avgRed, 
The overhead of performing full semantic analysis on the entire program is on median 19x, ranging between 3x and 67x. 
While the semantic analysis results at $\infty$ level represent the most precise analysis our technique achieves, it is quite expensive, and it even times-out for our largest program (e.g. {\tt space}). 
For example in the \Use{pn.theft} project the reduction achieved by $\semDataflowcia_{\infty}$ is 77\% but with a 64x overhead. 
This motivates the need for an incremental analysis, whose results can be obtained faster.

\Comment{
\input{anecdote-non-refactoring}

\myparabf{Non-Refactoring:} \Fix{Drop this to save space?}
In some cases $\semDataflowcia$ does not improve on the $\dataflowcia$ analysis.
\fig{}~\ref{fig:non-refactor} shows an example of a failed refactoring.
 The developer tries to refactor the code to use a variable rather than a constant and adds some parentheses \textit{as a matter of style} to the expression. 
 Unfortunately, parenthesizing the entire expression leaves the division operating over integers and performing truncation, and only converts the final result to a \codein{float}.   
 This was not the behavior before the change and this change breaks the program, introducing a bug, which was fixed by another developer 5 months later (by removing the parentheses). 
$\semimp$ finds that this change impacts many other portions of code and hence warns the developer that the change is not an actual refactoring. 
The developer may inspect the change more thoroughly or write some tests to ensure the change does not change the program in unintended ways.
}

\myparabf{Imprecision}
Our manual inspection of results reveals three broad classes for nodes classified as impacted:
(i)~nodes in syntactically changed procedures, (ii)~\SymDiff's inability to match loops as it relies on syntactic position in AST (this can be fixed by better matching heuristics), (iii)~SMACK represents all aliased addresses accessing a field using a single map; writing to one location destroys equivalences on the map variables (need more refined conditional equivalences~\cite{lahiri-condeq-tr10}).

\subsection{Incremental Analysis}
\label{sec:eval:incremental}

\tab~\ref{tab:results} shows the results of varying the bound on $k$ for the \anytimeAlgo.
The first iteration $\semDataflowcia_0$ corresponds to semantically analyzing only the syntactically-changed procedures; the second iteration $\semDataflowcia_1$ corresponds to analyzing the procedures at distance at most one from the syntactically changed procedures. 
The results show that even $\semDataflowcia_0$ provides benefits, pruning the impacted set by \Use{0.avg.red} on average. 
The overhead is reduced compared to the full analysis (\Use{0.avg.overhead}). 
Similarly, results for $\semDataflowcia_1$ show the analysis is effective. 
For the case of \Use{pn.theft} the improvement is from 61\% ($\semDataflowcia_0$) to 77\% ($\semDataflowcia_\infty$), at the cost of overhead increase from 8x to 64x.

\input{tables/space_table}

In general, we foresee that the anytime analysis is most useful for cases where it is prohibitive to run the full algorithm because of time constraints. 
This is best illustrated for the case of \texttt{space} (we used a timeout of one hour).
\tab~\ref{tab:space_bounds} shows the first four levels for \texttt{space} (two more iteration beyond the ones in \tab~\ref{tab:results});  performing the analysis incrementally is still valuable even upto $k$ = 3; the first iteration already provides big benefits on top of the non-semantic analysis, while the following iterations display a smooth improvement with each iteration. 
We believe this highlights the benefits of our anytime algorithm, giving the user control over the tradeoff between precision and analysis-time. 
\Comment{
\input{figures/depth_comparison}

To gain better insight on how the depth of our analysis, i.e., the number of iterations run, influences the impacted set we highlight how the impacted set is distributed across the program with respect to the syntactically changed procedure.
We show that the impacted statements are distributed across the program and how more iterations are able to prune impacted statements that are farther away from the syntactic change.
For illustration purposes, in \fig~\ref{fig:comparing_depth} we show the distribution of the impacted set, across the program procedures, by distance from the syntactically changed procedure; the x axis shows the distance from the syntactically changed procedure and the y axis shows the number of nodes impacted at the corresponding distance.  We show results for the first, second, and third iteration of the analysis. Blue lines show information for non-semantic analysis, while the red one shows results for the semantic analysis. 
For example, for the first iteration, corresponding to depth 0, there are 29 lines impacted at distance 0 from the change, i.e., in the syntactically changed procedure itself; 32 lines at distance 1 by the syntactic analysis and similarly for the semantic analysis.
The three figures convey the intuitive idea that as more iterations are performed, impacts from farther away from the change are pruned.
}

\subsection{Representative Examples}
\label{sec:anecdotes}

An inspection of the results indicates that the improvement in precision in \semDataflowcia comes from two fronts.
First, it compensates for the price we paid for soundness by considering entire procedures as source of impact. 
The semantic analysis\Comment{, although does not reduce the impacted set for the $\changedProcs{}$ it} reduces the impacts for callers and callees transitively.
Second, the reduction in impact happens from refactorings that a pure dataflow analysis cannot consider. 
We show a few interesting patterns that we discovered while applying the tool (for brevity we only describe the change briefly). 

\input{anecdote-constant-inline}

\input{anecdote-for-while}

\input{anecdote-go-to}

\myparabf{Variable Extraction} 
\fig{}~\ref{fig:extract-refactor} shows a  refactoring to extract a constant to a variable. A non-semantic technique will create impacts in \codein{term\_move\_to} through the first argument, since it will not be able to find that the value flowing into the first argument is the same in both versions and in all executions. Our $\semimp$ technique will successfully prove the mutual precondition necessary to show the equality in both versions, and hence cut impacts that would propagate through the first argument.

\myparabf{Loop Refactoring}
\fig{}~\ref{fig:loop-refactor} shows a change from a while loop to a for loop.
Input-output equivalence checking would not prevent the impact of the argument \codein{c} to the callee inside the loop (nor would dataflow analysis). 
Remember that we extract loops as tail recursive procedures.
%
%

\myparabf{Control-Flow Equivalence} 
\fig{}~\ref{fig:goto-refactor} shows a change to replace a \codein{goto} with \codein{return} statements.
This is a change in the project \Use{pn.tinyvm}. The \codein{goto} statements were all redirecting control-flow to a \codein{return} statement, so the developer replaced the \codein{goto} with the target \codein{return} statement. 


%% file: tables/projects_summary.tex

\begin{table}
\begin{center}
\footnotesize
\begin{tabular}{|l|r|r|r|r|r|}

\hline
\multicolumn{1}{|c|}{Project} &\multicolumn{1}{|c|}{\# Version}& \multicolumn{2}{|c|}{SLOC} & \multicolumn{2}{|c|}{LOC Changed}\\
\multicolumn{1}{|c|}{Name} & \multicolumn{1}{|c|}{Pairs} & \multicolumn{1}{|c|}{min} & \multicolumn{1}{|c|}{max} & \multicolumn{1}{|c|}{min} & \multicolumn{1}{|c|}{max}\\
\hline
\Use{pn.flingfd} & \Use{vn.flingfd} & \Use{sloc.min.flingfd} & \Use{sloc.max.flingfd} & \Use{cng.min.flingfd} & \Use{cng.max.flingfd}\\
\Use{pn.histo} & \Use{vn.histo} & \Use{sloc.min.histo} & \Use{sloc.max.histo} & \Use{cng.min.histo} & \Use{cng.max.histo}\\
\Use{pn.mdp} & \Use{vn.mdp} & \Use{sloc.min.mdp} & \Use{sloc.max.mdp} & \Use{cng.min.mdp} & \Use{cng.max.mdp}\\
\Use{pn.theft} & \Use{vn.theft} & \Use{sloc.min.theft} & \Use{sloc.max.theft} & \Use{cng.min.theft} & \Use{cng.max.tinyvm}\\
\Use{pn.tinyvm} & \Use{vn.tinyvm} & \Use{sloc.min.tinyvm} & \Use{sloc.max.tinyvm} & \Use{cng.min.tinyvm} & \Use{cng.max.tinyvm}\\
\hline\hline
\Use{pn.pt} & \Use{vn.pt} & \Use{sloc.min.pt} & \Use{sloc.max.pt} & \Use{cng.min.pt} & \Use{cng.max.pt}\\
\Use{pn.pt2} & \Use{vn.pt2} & \Use{sloc.min.pt2} & \Use{sloc.max.pt2} & \Use{cng.min.pt2} & \Use{cng.max.pt2}\\
\Use{pn.replace} & \Use{vn.replace} & \Use{sloc.min.replace} & \Use{sloc.max.replace} & \Use{cng.min.replace} & \Use{cng.max.replace}\\
\Use{pn.schedule} & \Use{vn.schedule} & \Use{sloc.min.schedule} & \Use{sloc.max.schedule} & \Use{cng.min.schedule} & \Use{cng.max.schedule}\\
\Use{pn.space} & \Use{vn.space} & \Use{sloc.min.space} & \Use{sloc.max.space} & \Use{cng.min.space} & \Use{cng.max.space}\\
\Use{pn.tcas} & \Use{vn.tcas} & \Use{sloc.min.tcas} & \Use{sloc.max.tcas} & \Use{cng.min.tcas} & \Use{cng.max.tcas}\\
\Use{pn.tot_info} & \Use{vn.tot_info} & \Use{sloc.min.tot_info} & \Use{sloc.max.tot_info} & \Use{cng.min.tot_info} & \Use{cng.max.tot_info}\\
\hline
\end{tabular}
\end{center}
\caption{Summary of projects used as evaluation subjects}
\label{tab:proj_summary}
\vspace{\vsp}
\vspace{\vsp}
\vspace{\vsp}
\end{table}

%% file: tables/results_dac_linear.tex

\begin{table*}
\begin{center}
\scriptsize
\begin{tabular}{|l||r|r|r||r|r|r|r||r|r|r|r||r|r|r|r|}
\hline
\multicolumn{1}{|c||}{Project} &\multicolumn{3}{|c||}{\dataflowcia}& \multicolumn{4}{|c||}{$\semDataflowcia_0$}& \multicolumn{4}{|c||}{$\semDataflowcia_1$}& \multicolumn{4}{|c|}{$\semDataflowcia_{\infty}$}\\
\multicolumn{1}{|c||}{Name} & \multicolumn{1}{|c|}{min} & \multicolumn{1}{|c|}{max}& \multicolumn{1}{|c||}{Time} &  \multicolumn{1}{|c|}{min} & \multicolumn{1}{|c|}{max} &\multicolumn{1}{|c|}{Red}& \multicolumn{1}{|c||}{Time} &  \multicolumn{1}{|c|}{min} & \multicolumn{1}{|c|}{max} &\multicolumn{1}{|c|}{Red}& \multicolumn{1}{|c||}{Time}&  \multicolumn{1}{|c|}{min} & \multicolumn{1}{|c|}{max} &\multicolumn{1}{|c|}{Red}& \multicolumn{1}{|c|}{Time}\\
\hline
\Use{pn.flingfd}&\Use{0.dcf.min.flingfd}&\Use{0.dcf.max.flingfd}&\Use{0.dcf.time.flingfd}&\Use{0.sem.min.flingfd}&\Use{0.sem.max.flingfd}&\Use{0.avg.red.flingfd}&\Use{0.sem.time.flingfd}&\Use{1.sem.min.flingfd}&\Use{1.sem.max.flingfd}&\Use{1.avg.red.flingfd}&\Use{1.sem.time.flingfd}&\Use{inf.sem.min.flingfd}&\Use{inf.sem.max.flingfd}&\Use{inf.avg.red.flingfd}&\Use{inf.sem.time.flingfd}\\

\Use{pn.histo}&\Use{0.dcf.min.histo}&\Use{0.dcf.max.histo}&\Use{0.dcf.time.histo}&\Use{0.sem.min.histo}&\Use{0.sem.max.histo}&\Use{0.avg.red.histo}&\Use{0.sem.time.histo}&\Use{1.sem.min.histo}&\Use{1.sem.max.histo}&\Use{1.avg.red.histo}&\Use{1.sem.time.histo}&\Use{inf.sem.min.histo}&\Use{inf.sem.max.histo}&\Use{inf.avg.red.histo}&\Use{inf.sem.time.histo}\\

\Use{pn.mdp}&\Use{0.dcf.min.mdp}&\Use{0.dcf.max.mdp}&\Use{0.dcf.time.mdp}&\Use{0.sem.min.mdp}&\Use{0.sem.max.mdp}&\Use{0.avg.red.mdp}&\Use{0.sem.time.mdp}&\Use{1.sem.min.mdp}&\Use{1.sem.max.mdp}&\Use{1.avg.red.mdp}&\Use{1.sem.time.mdp}&\Use{inf.sem.min.mdp}&\Use{inf.sem.max.mdp}&\Use{inf.avg.red.mdp}&\Use{inf.sem.time.mdp}\\

\Use{pn.tinyvm}&\Use{0.dcf.min.tinyvm}&\Use{0.dcf.max.tinyvm}&\Use{0.dcf.time.tinyvm}&\Use{0.sem.min.tinyvm}&\Use{0.sem.max.tinyvm}&\Use{0.avg.red.tinyvm}&\Use{0.sem.time.tinyvm}&\Use{1.sem.min.tinyvm}&\Use{1.sem.max.tinyvm}&\Use{1.avg.red.tinyvm}&\Use{1.sem.time.tinyvm}&\Use{inf.sem.min.tinyvm}&\Use{inf.sem.max.tinyvm}&\Use{inf.avg.red.tinyvm}&\Use{inf.sem.time.tinyvm}\\

\Use{pn.theft}&\Use{0.dcf.min.theft}&\Use{0.dcf.max.theft}&\Use{0.dcf.time.theft}&\Use{0.sem.min.theft}&\Use{0.sem.max.theft}&\Use{0.avg.red.theft}&\Use{0.sem.time.theft}&\Use{1.sem.min.theft}&\Use{1.sem.max.theft}&\Use{1.avg.red.theft}&\Use{1.sem.time.theft}&\Use{inf.sem.min.theft}&\Use{inf.sem.max.theft}&\Use{inf.avg.red.theft}&\Use{inf.sem.time.theft}\\

\hline\hline

\Use{pn.pt}&\Use{0.dcf.min.print_tokens}&\Use{0.dcf.max.print_tokens}&\Use{0.dcf.time.print_tokens}&\Use{0.sem.min.print_tokens}&\Use{0.sem.max.print_tokens}&\Use{0.avg.red.print_tokens}&\Use{0.sem.time.print_tokens}&\Use{1.sem.min.print_tokens}&\Use{1.sem.max.print_tokens}&\Use{1.avg.red.print_tokens}&\Use{1.sem.time.print_tokens}&\Use{inf.sem.min.print_tokens}&\Use{inf.sem.max.print_tokens}&\Use{inf.avg.red.print_tokens}&\Use{inf.sem.time.print_tokens}\\

\Use{pn.pt2}&\Use{0.dcf.min.print_tokens2}&\Use{0.dcf.max.print_tokens2}&\Use{0.dcf.time.print_tokens2}&\Use{0.sem.min.print_tokens2}&\Use{0.sem.max.print_tokens2}&\Use{0.avg.red.print_tokens2}&\Use{0.sem.time.print_tokens2}&\Use{1.sem.min.print_tokens2}&\Use{1.sem.max.print_tokens2}&\Use{1.avg.red.print_tokens2}&\Use{1.sem.time.print_tokens2}&\Use{inf.sem.min.print_tokens2}&\Use{inf.sem.max.print_tokens2}&\Use{inf.avg.red.print_tokens2}&\Use{inf.sem.time.print_tokens2}\\

\Use{pn.replace}&\Use{0.dcf.min.replace}&\Use{0.dcf.max.replace}&\Use{0.dcf.time.replace}&\Use{0.sem.min.replace}&\Use{0.sem.max.replace}&\Use{0.avg.red.replace}&\Use{0.sem.time.replace}&\Use{1.sem.min.replace}&\Use{1.sem.max.replace}&\Use{1.avg.red.replace}&\Use{1.sem.time.replace}&\Use{inf.sem.min.replace}&\Use{inf.sem.max.replace}&\Use{inf.avg.red.replace}&\Use{inf.sem.time.replace}\\

\Use{pn.schedule}&\Use{0.dcf.min.schedule}&\Use{0.dcf.max.schedule}&\Use{0.dcf.time.schedule}&\Use{0.sem.min.schedule}&\Use{0.sem.max.schedule}&\Use{0.avg.red.schedule}&\Use{0.sem.time.schedule}&\Use{1.sem.min.schedule}&\Use{1.sem.max.schedule}&\Use{1.avg.red.schedule}&\Use{1.sem.time.schedule}&\Use{inf.sem.min.schedule}&\Use{inf.sem.max.schedule}&\Use{inf.avg.red.schedule}&\Use{inf.sem.time.schedule}\\

\Use{pn.space}&\Use{0.dcf.min.space}&\Use{0.dcf.max.space}&\Use{0.dcf.time.space}&\Use{0.sem.min.space}&\Use{0.sem.max.space}&\Use{0.avg.red.space}&\Use{0.sem.time.space}&\Use{1.sem.min.space}&\Use{1.sem.max.space}&\Use{1.avg.red.space}&\Use{1.sem.time.space}&\Use{inf.sem.min.space}&\Use{inf.sem.max.space}&\Use{inf.avg.red.space}&\Use{inf.sem.time.space}\\

\Use{pn.tcas}&\Use{0.dcf.min.tcas}&\Use{0.dcf.max.tcas}&\Use{0.dcf.time.tcas}&\Use{0.sem.min.tcas}&\Use{0.sem.max.tcas}&\Use{0.avg.red.tcas}&\Use{0.sem.time.tcas}&\Use{1.sem.min.tcas}&\Use{1.sem.max.tcas}&\Use{1.avg.red.tcas}&\Use{1.sem.time.tcas}&\Use{inf.sem.min.tcas}&\Use{inf.sem.max.tcas}&\Use{inf.avg.red.tcas}&\Use{inf.sem.time.tcas}\\

\Use{pn.tot_info}&\Use{0.dcf.min.tot_info}&\Use{0.dcf.max.tot_info}&\Use{0.dcf.time.tot_info}&\Use{0.sem.min.tot_info}&\Use{0.sem.max.tot_info}&\Use{0.avg.red.tot_info}&\Use{0.sem.time.tot_info}&\Use{1.sem.min.tot_info}&\Use{1.sem.max.tot_info}&\Use{1.avg.red.tot_info}&\Use{1.sem.time.tot_info}&\Use{inf.sem.min.tot_info}&\Use{inf.sem.max.tot_info}&\Use{inf.avg.red.tot_info}&\Use{inf.sem.time.tot_info}\\

\hline
\end{tabular}
\end{center}
\caption{Analysis results for different levels of precision. Time in seconds. (timeout = 1 hour)}
\label{tab:results}
\vspace{\vsp}\vspace{\vsp}
\vspace{\vsp}
\end{table*}

%% file: anecdote-non-refactoring.tex

\begin{figure}
{\footnotesize{
\begin{lstlisting}[style=C-github, numbers=none]
...
@d@-if((float) ++htab->num_nodes / htab->size  > @df@0.7@df@)@d@
@a@+if((float)@af@(@af@++htab->num_nodes / htab->size@af@)@af@ > @af@HtabLoadFactor@af@)@a@
...
\end{lstlisting}
}}
\caption{Change illustrating a buggy refactoring in \Use{pn.tinyvm} commit \textit{716936}}
\label{fig:non-refactor}
\vspace{\vsp}
\end{figure}

%% file: tables/space_table.tex

\begin{table}
\begin{center}
\footnotesize
\begin{tabular}{|l|r|r|r|r|}

\hline
Analysis&Min&Max&Reduction&Time\\
\hline
\dataflowcia&\Use{0.dcf.min.space}&\Use{0.dcf.max.space}&\na&\Use{0.dcf.time.space}\\
$\semDataflowcia_0$&\Use{0.sem.min.space}&\Use{0.sem.max.space}&\Use{0.avg.red.space}&\Use{0.sem.time.space}\\
$\semDataflowcia_1$&\Use{1.sem.min.space}&\Use{1.sem.max.space}&\Use{1.avg.red.space}&\Use{1.sem.time.space}\\
$\semDataflowcia_2$&\Use{2.sem.min.space}&\Use{2.sem.max.space}&\Use{2.avg.red.space}&\Use{2.sem.time.space}\\
$\semDataflowcia_3$&\Use{3.sem.min.space}&\Use{3.sem.max.space}&\Use{3.avg.red.space}&\Use{3.sem.time.space}\\
$\semDataflowcia_{\infty}$&\Use{inf.sem.min.space}&\Use{inf.sem.max.space}&\Use{inf.avg.red.space}&\Use{inf.sem.time.space}\\

\hline
\end{tabular}
\end{center}
\caption{Analysis results for space}
\label{tab:space_bounds}
\vspace{\vsp}
\vspace{\vsp}
\end{table}

%% file: figures/depth_comparison.tex

\begin{figure*}[ht!]

  \begin{tabular}{@{}c@{} @{}c@{} @{}c@{}}
      \includegraphics[width=.32\linewidth]{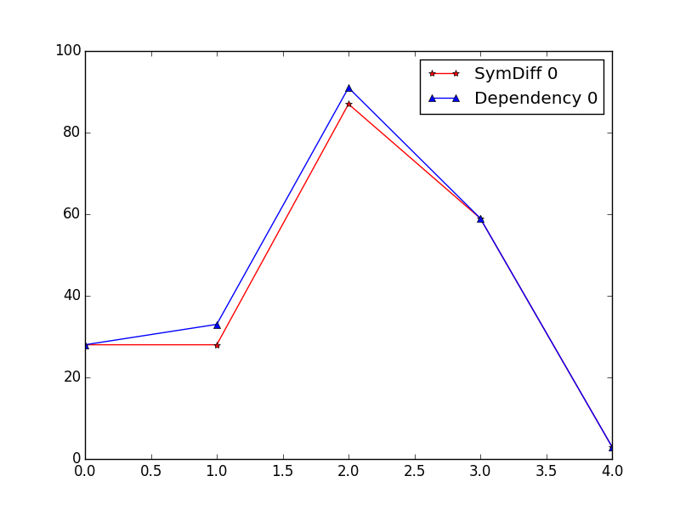} 

 &    \includegraphics[width=.32\linewidth]{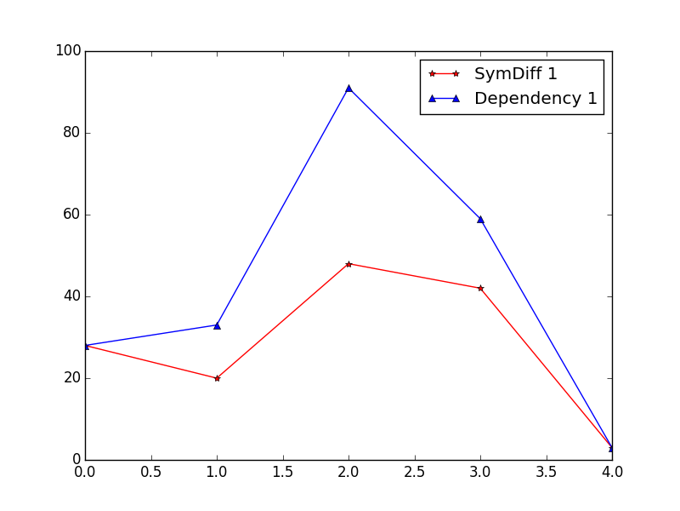} 
 
 &    \includegraphics[width=.32\linewidth]{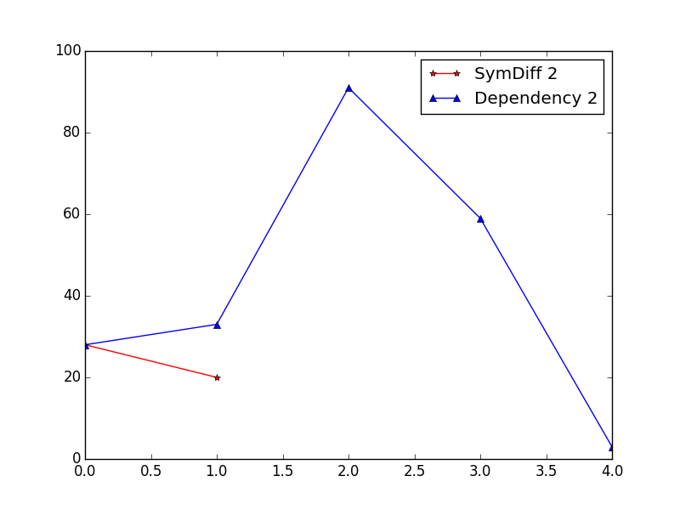}\\ 
(0)&(1)&(2)\\
  \end{tabular}

\caption{Comparing results for different analysis depths for the \Use{pn.tinyvm} project}
\label{fig:comparing_depth}
\vspace{\vsp}
\end{figure*}

%% file: anecdote-constant-inline.tex

\begin{figure}
\vspace{\vsp}
\begin{minipage}[t]{0.3\textwidth}
\begin{lstlisting}[style=C-github, numbers=none]
void draw_histogram(int data[], int len) {
  ...
@a@+ int xbarw = 5;@a@
  ...
  while (y--) {
@d@-   term_move_to(x * @df@5@df@ + xpad + 3,@d@ 
@a@+   term_move_to(x * @af@xbarw@af@ + xpad + 3,@a@
        y - 1 + h + ypad);
      ...
    }  
}
\end{lstlisting}

\end{minipage}
\vspace{\vsp}
\caption{Change illustrating an extract constant to variable in \Use{pn.histo} commit \textit{c723a4}}
\label{fig:extract-refactor}
\vspace{\vsp}
\end{figure}

%% file: anecdote-for-while.tex

\begin{figure}[tb]
\vspace{\vsp}
\vspace{\vsp}
\begin{minipage}[t]{0.2\textwidth}
\begin{lstlisting}[style=C-github, numbers=none]
@d@-while (*c) {@d@
@a@+for (;*c;c++) {@a@
  ...
  wprintw(window, "%c",  *c);
@d@- c++;@d@
}
\end{lstlisting}
\end{minipage}
\caption{Change illustrating a loop conversion in \Use{pn.mdp} commit \textit{00c2ad}}
\label{fig:loop-refactor}
\vspace{\vsp}
\end{figure}

%% file: anecdote-go-to.tex

\begin{figure}[tb]
\vspace{\vsp}
\begin{minipage}[t]{0.4\textwidth}
\begin{lstlisting}[style=C-github, numbers=none]
...
@d@- if (!strend || !strbegin) @df@goto pp_ret@df@;@d@
@a@+ if (!strend || !strbegin) @af@return 0@af@;@a@

  if (!pFile) {
    ...
@d@-   @df@goto pp_ret@df@;@d@
@a@+   @af@return 0@af@;@a@
  }
  ...
@d@- @df@pp_ret: @df@return 0;@d@
@a@+ return 0;@a@
}
\end{lstlisting}
\end{minipage}
\vspace{\vsp}
\caption{Change illustrating a goto-elimination refactoring in \Use{pn.tinyvm} commit \textit{378cc6}}
\label{fig:goto-refactor}
\vspace{\vsp}
\end{figure}

%% file: related.tex

\section{Related Work}

Our work is closely related to work on change-impact analysis, regression verification, regression test generation, and relational analysis. 

\myparabf{Change impact analysis} 
Change Impact Analysis has been widely explored in static and dynamic program analysis context~\cite{lehnert2011review, haipeng2015survey, ren2004chianti,ryder2001change, law2003whole}. 
Most previous works perform the analysis at a coarse-grain level (classes and types) to retain soundness of analysis~\cite{apiwattanapong2005icse, apiwattanapong2004differencing, orso2003leveraging,le2014patchVerificationICSE} which can result in coarse results.
JDiff~\cite{apiwattanapong2004differencing} addresses some of the challenges of performing both a diff and computing a mapping between two programs in the context of Java object-oriented programs.
Other techniques resort to dynamic information to recover from the overly-conservative dataflow analysis~\cite{orso2003leveraging,apiwattanapong2005icse}.
Our goal is to improve the precision of CIA analysis by making it change-semantics aware using statically computed equivalence relations without sacrificing soundness. 

\myparabf{Regression verification}
Regression verification~\cite{godlin-dac09,Person2008DSE} and its implementations~\cite{lahiri-cav12} aim at proving summary equivalence interprocedurally, but does not help with the CIA directly as shown in \S~\ref{sec:overview}.
The work by Bakes et al.~\cite{backes2013regression} improves traditional equivalence checking by pruning paths not impacted by changes. 
The approach is non-modular (does not summarize callees), bounded (unrolls loops and recursion), and does not seek to improve the underlying change-impact analysis. 
The technique leverages CIA to avoid performing equivalence checking on non-impacted procedures (computed by standard dataflow analysis).
These approaches are useful for equivalence-preserving changes; when the changes are non-equivalent they do not provide meaningful help for reducing code review or testing efforts.
Our approach, on the other hand, refines the CIA and can be used in code review and regression testing. 
Besides, our approach retains modularity and is sound in the presence of loops and recursion. 
We leverage the product construction in \SymDiff~\cite{lahiri-fse13} that has been used for {\it differential assertion checking} (checking if an assertion fails more often after a change); however this work is limited as it requires the presence of assertions in the program. 
Our approach can also use other product construction techniques and relational invariant inference techniques as an off-the-shelf solver~\cite{barthe2011relational,barthe2013beyond,carbinpldi}.

\myparabf{Regression testing}
Person et al. using change directed symbolic execution to generate regression tests~\cite{Person2011DISE}.
Our technique can be used to prune the space of statements for which regression tests need to be generated.
In addition, there is research on relational verification using a product construction~\cite{sas-PartushY13,barthe2011relational,barthe2013beyond,benton-popl06}, but most approaches are not automated and do not consider changes across procedure calls. 



%% file: conclusions.tex

\section{Conclusions}

In this work, we have formalized and demonstrated how to leverage equivalence relations to improve the precision of change-impact analysis and provide a scalability-precision knob with \anytimeAlgo, which is crucial for applying such analyses to large projects.
Our work brings together program verification techniques (namely relational invariant generation) to improve the precision of a core software engineering task, and can go a long way in providing the benefits of deep semantic reasoning to average developers. 